\documentclass[12pt,onecolumn, peerreview, draftcls, letterpaper]{IEEEtran}

\usepackage[usenames]{color}
\usepackage{url}
\usepackage{cite}      

\usepackage{graphicx}  

\usepackage{epsf}
\usepackage{epsfig}

\usepackage{subfigure} 

\usepackage{url}       

\usepackage{amsmath}   

\usepackage{lscape}

\renewcommand{\baselinestretch}{2}

\newcommand{\sizecorr}[1]{\makebox[0cm]{\phantom{$\displaystyle #1$}}}

\begin{document}

\title{Technical Report: Finger Replacement Schemes for {RAKE} Receivers in the Soft Handover Region with Multiple BSs over I.N.D. Fading Channels}

\author{
Sung~Sik~Nam, Seyeong Choi,
Sung Ho Cho, and Mohamed-Slim~Alouini
\thanks{S.~S.~Nam and S.~H.~Cho are with Hanyang University, Korea.  S. Choi is with Wonkwang University, Korea.  M.-S. Alouini is with KAUST, Sauid Arabia.}
}

\markboth{S.S. Nam \MakeLowercase{\textit{et al.}}: Finger Replacement over I.N.D. }{Shell \MakeLowercase{\textit{et al.}}: Finger Replacement over I.N.D. }

\maketitle

\begin{abstract}
A new finger replacement technique which is applicable for RAKE receivers in the soft handover (SHO) region has been proposed and studied in~\cite{kn:S_Choi_2008_4, kn:S_Choi_2008_1} under the ideal assumption that the fading is both independent and identically distributed  from path to path. To supplement our previous work, we present a general comprehensive framework for the performance assessment of the proposed finger replacement schemes operating over  independent and non-identically distributed (i.n.d.) faded paths. To accomplish this object, we derive new closed-form expressions for the target key statistics which are composed of i.n.d. exponential random variables. With these new expressions, the performance analysis of various wireless communication systems over more practical environments can be possible.
\end{abstract}

\section{Introduction} \label{sec_1}
Multipath fading is an unavoidable physical phenomenon that affects
considerably the performance of wideband wireless communication
systems. While usually viewed as a deteriorating factor, multipath
fading can also be exploited to improve the performance by using
RAKE-type receivers. However, in the soft handover (SHO) region, due
to the limited number of fingers in the mobile unit, we are faced
with a problem of how to judiciously select a subset of paths for
the RAKE reception to achieve the required performance.

Finger replacement techniques for RAKE reception in
the SHO region have been proposed and analyzed over
independent and identical distributed (i.i.d.) fading environments
with two base stations (BSs) in ~\cite{kn:S_Choi_2008_4} which was
extended to the case of multiple BSs in~\cite{kn:S_Choi_2008_1}. The
proposed schemes in~\cite{kn:S_Choi_2008_1}, as shown in
Fig.~\ref{example_2}, are basically based on the block comparison
among groups of resolvable paths from different BSs and lead to the
reduction of complexity while offering commensurate performance in
comparison with previously proposed schemes
in~\cite{kn:S_Choi_2008_2, kn:S_Choi_2008_3}. However, in practice,
the i.i.d. fading scenario on the diversity paths is not always
realistic due to, for example, the different adjacent multipath
routes with the same path loss and the resulting unbalance among paths.
Although this non-identical consideration is important from a practical
standpoint, \cite{kn:S_Choi_2008_1} was able to investigate the effect of the non-uniform
power delay profile of the finger replacement schemes only with
computer simulations due to the high complexity of the analysis.
Note that the applied method in \cite{kn:MS_GSC} to derive the required key statistics for i.i.d fading assumptions can not be directly adopted to the case of independent and non-identically distributed (i.n.d.) fading environments. The major difficulties lie in deriving the target statistics with non-identical parameters.

With this observation in mind, we mathematically attack these main difficulties in this report. More specifically, we address the key mathematical formalism which are the statistics of partial sums and the two-dimensional joint statistics of partial sums of the i.n.d. ordered random variables (RVs) for the accurate performance analysis of the finger replacement scheme with non-identical parameters.
The rest of this report is
organized as follows. In Section~\ref{sec_2}, we present the system
models as well as the mode of operation of the finger replacement scheme under
consideration and provide the results of a general comprehensive framework for the outage performance based on statistical results over
i.n.d. fading channels. We then provide in Section~\ref{sec_3} some closed-form expressions of the required key statistics. Finally, Section~\ref{conc}
provides some concluding remarks.

\section{System Models and Performance Measures}  \label{sec_2}
Among the path scanning schemes proposed in~\cite{kn:S_Choi_2008_1},
we consider the full scanning method. With this method, if the combined output signal-to-noise ratio (SNR) of
current assigned fingers is greater than a certain target SNR, a
one-way SHO is used and no finger replacement is needed. Otherwise,
the receiver attempts a two-way SHO by starting to scan additional
paths from the serving BS as well as all the target BSs.

We assume that $L$ BSs are active, and there are a total of
${N_{\left( L \right)}}$ resolvable paths, where ${N_{\left( L
\right)}} = \sum\limits_{n = 1}^L {{N_n}}$ and $N_n$  is the number
of resolvable paths from the $n$-th BS. In the SHO region, only
$N_c$  out of  ${N_{\left( n \right)}}$ $\left(1 \le n \le L\right)$
paths are used for RAKE reception. Without loss of generality, let
$N_1$ be the number of resolvable paths from the serving BS and
$N_2, N_3, \ldots ,N_L$ be those from the target BSs. In the SHO
region, the receiver is assumed at first to rely only on $N_1$
resolvable paths and, as such, starts with ${N_c}/{N_1}$-generalized
selection Combining (GSC)~\cite{kn:alouini_wi_j3}. These schemes are
based on the comparison of blocks consisting of ${N_s}\left( { <
{N_c} < {N_n}} \right)$ paths from each BS.

Let $u_{i,n}$ ($i=1,2,\ldots,N_n$) be the $i$-th order statistics
out of $N_n$ SNRs of paths from the $n$-th BS by arranging $N_n$
$(L\ge2)$ nonnegative i.n.d. RVs, $\left\{ {\gamma_{j,n} }
\right\}_{j = 1}^{N_n}$, at $n$-th BS, where $\gamma_{j,n}$ is the
the SNR of the $j$-th path from $n$-th BS, in decreasing order of
magnitude such that $u_{1,n} \ge u_{2,n} \ge \cdots \ge
u_{{N_n},n}$. If we let
\begin{equation} \small \label{eq:1}
Y = \sum\limits_{i = 1}^{{N_c} - {N_{s}}} {{u_{i,1}}}
\end{equation}
\begin{equation} \small  \label{eq:2}
{W_n} = \left\{ \begin{array}{l}
 \sum\limits_{i = {N_c} - {N_{s}} + 1}^{{N_c}} {{u_{i,n}}}, \,\,\quad \quad n = 1 \\
 \sum\limits_{i = 1}^{{N_{s}}} {{u_{i,n}}},  \quad \quad \quad \quad \quad \,\;\,n = 2, \ldots ,L , \\
 \end{array} \right.
\end{equation}
then the received output SNR after GSC is given by  $Y + {W_1}$. At
the beginning of every time slot, the receiver compares the GSC
output SNR, $Y + {W_1}$, with a certain target SNR. If $Y + {W_1}$ is
greater than or equal to the target SNR, a one-way SHO is used and
no finger replacement is needed. On the other hand, whenever $Y +
{W_1}$  falls below the target SNR, the receiver attempts a two-way
SHO by starting to scan additional paths from the target BSs.

To study the performance of the finger replacement scheme for i.n.d. fading
assumptions, we look into the outage performance. Based on the
mode of operation in Section \cite[II-B]{kn:S_Choi_2008_1}, an
overall outage probability is declared when
the final combined SNR, $\gamma_F$, falls below a predetermined
threshold, $x$, as
\begin{equation} \small \label{eq:3}
{F_{{\gamma _F}}}\left( x
\right) = \Pr \left[ {{\gamma _F} < x} \right]
\end{equation}
where
\begin{equation} \small\label{eq:4}
{\gamma _F} = \left\{ \begin{array}{lr}
 Y + {W_1}, &Y + {W_1} \ge {\gamma _T} \\
 Y + \max \left\{ {{W_1},{W_2}, \cdots ,{W_L}} \right\}, &Y + {W_1} < {\gamma _T}. \\
 \end{array} \right.
\end{equation}
Considering two cases that  i) the final combined SNR is greater than or equal to the
target SNR, $\gamma_T$, (i.e., $x \ge {\gamma _T}$) and ii) the
final combined SNR falls below the target SNR, (i.e., $0 < x <
{\gamma _T}$), separately,  we can
rewrite (\ref{eq:3}) as
\begin{equation} \small \label{eq:7}
{F_{{\gamma _F}}}\left( x \right) = \left\{ \begin{array}{ll}
 \Pr \left[ {Y + \max \left\{ {{W_1},{W_2}, \cdots ,{W_L}} \right\} < x} \right], & 0 < x < {\gamma _T} \\
 \Pr \left[ {{\gamma _T} \le Y + {W_1} < x} \right] \\
  + \Pr \left[ {Y + {W_1} < {\gamma _T},{\gamma _T} \le Y + \max \left\{ {{W_1},{W_2}, \cdots ,{W_L}} \right\} < x} \right],& x \ge {\gamma _T}.\\
 \end{array} \right.
\end{equation}
The detailed derivation is presented in the Appendix~\ref{appendix_1}.

With (\ref{eq:7}), we now need to investigate the following three probabilities,
a) $\Pr \big[ {{\gamma _T} \!\le \!Y\! + \!{W_1}\! < \!x} \big]$, b) $\Pr \big[ Y \!+\! {W_1} \!<\! {\gamma _T},{\gamma _T} \!\le\! Y\! + \!\max \left\{\! {W_1},\!{W_2}, \!\cdots ,\!{W_L}\! \right\} \!< x \big]$, and c) $ \Pr \big[ Y \!+ \!\max \{\! W_1, \!W_2, \!\cdots , \!W_L\! \} \!< x \big]$.
Note that the major
difficulty in the analysis is to derive the required key
statistics of ordered RVs. In~\cite{kn:S_Choi_2008_4} and \cite{kn:S_Choi_2008_1}, the required statistics were obtained by
applying the conditional probability density function (PDF) based
approach proposed in \cite{kn:MS_GSC} which is only valid for an assumption of i.i.d.
fading from path to path. However, in this report, our concern is that the
average SNR of each path (or branch) is different, which means more
practical channel models. For i.n.d. consideration unlike the
i.i.d. case, we need to consider realistic frequency selective
channels which have non-uniform delay profile, for example,
exponentially decaying power delay profile, and to deal with order statistics of
i.n.d. RVs. As results, the proposed method in \cite{kn:MS_GSC} can
not be directly adopted in case of i.n.d. fading environments here.

Recently, a unified framework to determine the joint statistics of
partial sums of ordered i.i.d. RVs has been introduced
in~\cite{kn:unified_approach}. With this proposed approach, the
required key statistics of any partial sums of ordered RVs can be
obtained systematically in terms of the moment generating function
(MGF) and the PDF. The extension of the mathematical approach
proposed in \cite{kn:unified_approach} to i.n.d. fading channels can
be found in \cite{kn:sungsiknam2013_ISIT, kn:IND_MGF_sungsiknam_1}. With the help of
\cite{kn:unified_approach, kn:sungsiknam2013_ISIT, kn:IND_MGF_sungsiknam_1}, the
required key statistics to investigate the outage probability
in~(\ref{eq:7}) over i.n.d. fading channels can be obtained.

Note that based on the mode of operation, $Y$ and $W_1$   are correlated while $W_n$ (for $n = 2,
\ldots ,L$) is independent of $Y$.
Hence, by adopting the
proposed approach in \cite{kn:sungsiknam2013_ISIT,  kn:IND_MGF_sungsiknam_1} instead of applying \cite{kn:MS_GSC},
required key statistics in (\ref{eq:7}) can be evaluated as
\begin{enumerate}
\item [a)] For $\Pr \left[ {{\gamma _T} \le Y + {W_1} < x} \right]$,
\begin{equation} \small \label{eq:8}
\Pr \left[ {{\gamma _T} \le Y + {W_1} < x} \right] = {F_{Y + {W_1}}}\left( x \right) - {F_{Y + {W_1}}}\left( {{\gamma _T}} \right).
\end{equation}
\item [b)] For $\Pr \left[ {Y + \max \left\{ {{W_1},{W_2}, \cdots ,{W_L}} \right\} < x} \right]$,
\begin{equation} \small  \label{eq:9}
\begin{aligned}
&\Pr \left[ {Y + \max \left\{ {{W_1},{W_2}, \cdots ,{W_L}} \right\} < x} \right] \\
&= \int_0^x {\int_0^{x - y} {{f_{Y,{W_1}}}\left( {y,{w_1}} \right)\int_0^{x - y} {{f_{{W_2}}}\left( {{w_2}} \right)d{w_2} \cdots \int_0^{x - y} {{f_{{W_L}}}\left( {{w_L}} \right)} d{w_L}d{w_1}dy} } }  \\
&= \int_0^x {\int_0^{x - y} {{f_{Y,{W_1}}}\left( {y,{w_1}} \right)\prod\limits_{n = 2}^L {{F_{{W_n}}}\left( {x - y} \right)} d{w_1}dy} }.
\end{aligned}
\end{equation}
\item [c)] For $\Pr \left[ {Y + {W_1} < {\gamma _T},{\gamma _T} \le Y + \max \left\{ {{W_1},{W_2}, \cdots ,{W_L}} \right\} < x} \right]$,
\begin{equation} \small  \label{eq:10}
\begin{aligned}
&\Pr \left[ {Y + {W_1} < {\gamma _T},{\gamma _T} \le Y + \max \left\{ {{W_1},{W_2}, \cdots ,{W_L}} \right\} < x} \right] \\
&= \int_0^{{\gamma _T}} {\int_0^{{\gamma _T} - y} {{f_{Y,{W_1}}}\left( {y,{w_1}} \right)\int_0^{x - y} {{f_{{W_2}}}\left( {{w_2}} \right)d{w_2} \cdots \int_0^{x - y} {{f_{{W_L}}}\left( {{w_L}} \right)} d{w_L}d{w_1}dy} } }  \\
&= \int_0^{{\gamma _T}} {\int_0^{{\gamma _T} - y} {{f_{Y,{W_1}}}\left( {y,{w_1}} \right)\prod\limits_{n = 2}^L {{F_{{W_n}}}\left( {x - y} \right)} d{w_1}dy} }.
\end{aligned}
\end{equation}
\end{enumerate}

Similar to the identical case in~\cite{kn:S_Choi_2008_1}, it is also
very important to study the complexity of finger replacement schemes over
i.n.d. case by accurately quantifying the performance measures such
as the average number of path estimations, the average number of SNR
comparisons, and the SHO overhead, which are required during the SHO
process of these schemes over i.n.d. case. Note that with these
performance measures, a comprehensive investigation of the tradeoff
between complexity and performance over i.n.d. fading channels can
be feasible. These important design parameters can be evaluated by
directly applying the defined formulas presented
in~\cite{kn:S_Choi_2008_1} with the required key statistics
for i.n.d. ordered RVs which will be derived in this work. Hence,
based on the mathematical approach proposed
in~\cite{kn:unified_approach, kn:sungsiknam2013_ISIT,  kn:IND_MGF_sungsiknam_1}, we
here focus on the derivation of the following key statistics such as the
cumulative distribution function (CDF) of the ${N_c}/{N_1}$-GSC
output SNR, ${F_{Y + {W_1}}}\left( \cdot \right)$, the 2-dimensional joint PDF of
two adjacent partial sums, $Y$ and $W_1$, of order statistics,
${{f_{Y,{W_1}}}\left( \cdot,\cdot \right)}$, and the CDF of the sum
of the $N_s$ strongest paths from each target BS,
${{F_{{W_n}}}\left( \cdot \right)}$, (i.e., $2 \le n \le L$).

\section{Key Statistics}  \label{sec_3}
In this section, we introduce the key statistics which are
essential to solve Eqs. (\ref{eq:8}), (\ref{eq:9}), and
(\ref{eq:10}) in Sec.~\ref{sec_2}. More specifically, in these three
cases, only the best ${N_c}$ or $N_s$ among $N_n$ ($N_s \le {N_c}
\le {N_n}$) ordered RVs are involved in the partial sums. Thus,
based on the unified frame work in \cite{kn:unified_approach} and
the extended work for i.n.d. case in \cite{kn:sungsiknam2013_ISIT,  kn:IND_MGF_sungsiknam_1},
each key statistics for three cases can be derived by applying the special step approach based on the substituted groups instead of original groups for each cases (i.e., starting from
2-dimensional joint statistics, 4-dimensional joint statistics, and 2-dimensional joint statistics, respectively) as
\begin{enumerate}
  \item [1)] ${F_{Y + {W_1}}}\left( x \right)$:
If we let $Z'=Y + {W_1}$ where $Y+W_1=\sum\limits_{i=1}^{N_c}
{u_{i,1}}$ for convenience, then we can derive the target CDF of
$Z'$ with the 2-dimensional joint PDF of
$Z_1=\sum\limits_{i=1}^{N_c-1} {u_{i,1}}$ and $Z_2={u_{N_c,1}}$ as
\begin{equation} \small \label{eq:11}
\begin{aligned}
 {F_{Y + {W_1}}}\left( x \right) =& \int_0^x {{f_{Z'}}\left( z \right)dz}  \\
  =& \int_0^x {\int_0^{\frac{z}{{_{{N_c}}}}} {{f_{{Z_1},{Z_2}}}\left( {z - {z_2},{z_2}} \right)d{z_2}} dz} .
\end{aligned}
\end{equation}
  \item [2)] ${{f_{Y,{W_1}}}\left( x, y\right)}$:
In this case, we can derive the target 2-dimensional PDF of
$Y=\sum\limits_{i = 1}^{{N_c} - {N_{s}}} {{u_{i,1}}}$ and
$W_1=\sum\limits_{i = {N_c} - {N_{s}} + 1}^{{N_c}} {{u_{i,1}}}$ by
transferring the 4-dimensional joint PDF of $Z_1  = \sum\limits_{i =
1}^{{N_c-N_s} - 1} {u_{i,1} }$, $Z_2  = u_{N_c-N_s,1}$, $Z_3 =
\sum\limits_{i = {N_c-N_s} + 1}^{N_c  - 1} {u_{i,1} }$, and $Z_4  =
u_{N_c,1}$ with the help of a function of a marginal PDF
as
\begin{equation} \small \label{eq:12} {f_{Y,{W_1}}}\left( {x,y}
\right) = \int_0^{\frac{y}{{{N_s} }}}
{\int_{\frac{y}{{{N_s}}}}^{\frac{x}{{N_c-N_s}}}
{{f_{{Z_1},{Z_2},{Z_3},{Z_4}}}\left( {x - {z_2},{z_2},y -
{z_4},{z_4}} \right)d{z_2}d{z_4}} }.
\end{equation}
  \item [3)] ${{F_{{W_n}}}\left( x \right)}$, (i.e., $2 \le n \le L$):
Similar to case 1),  the target one-dimensional CDF of
$W_n=\sum\limits_{i=1}^{N_s} {u_{i,n}}$ with the 2-dimensional joint
PDF of $Z'_1=\sum\limits_{i=1}^{N_s-1} {u_{i,n}}$ and
$Z'_2={u_{N_s,n}}$ can be derived with the help of a function of a
marginal PDF as
\begin{equation} \small \label{eq:13}
 {F_{{W_n}}}\left( x \right) = \int_0^x {\int_0^{\frac{z}{{_{{N_s}}}}} {{f_{{Z'_1},{Z'_2}}}\left( {z - {z'_2},{z'_2}} \right)d{z'_2}} dz}.
\end{equation}
\end{enumerate}

The above novel generic results in (\ref{eq:11})-(\ref{eq:13}) are
quite general and can be applied for any RVs. In this report, we limit our
analysis to the i.n.d. RVs case with a common exponential PDF,
${p_{{i_{l,n}}}}\left( x \right) = \frac{1}{{{{\bar \gamma
}_{{i_{l,n}}}}}}\exp \left( { - \frac{x}{{{{\bar \gamma
}_{{i_{l,n}}}}}}} \right)$ and CDF, $P_{{i_{l,n}}}\left( x \right) =
1-\exp \left( { - \frac{x}{{{{\bar \gamma }_{{i_{l,n}}}}}}} \right)$
for $\gamma\ge 0$, respectively, where ${\bar \gamma }_{i_{l,n}}$ is
the average of the $l$-th RV at $n$-th BS. Then, we can obtain the
target statistics in a ready-to-use form for i.n.d.
exponential RVs cases given as the following subsection. The
detailed derivations are presented in the Appendix~\ref{appendix_2} and \ref{appendix_3}.

\subsection{CDF of the ${N_c}/{N_1}$-GSC Output SNR over i.n.d. Rayleigh Fading, ${F_{Y + {W_1}}}\left( x \right)$}
\begin{equation} \scriptsize \label{eq:14}
\!\!\!\begin{aligned}
{F_{Y + {W_1}}}\left( x \right) &= \!\sum\limits_{{i_{{N_c,1}}} = 1}^{N_1}\! {\frac{1}{{{{\bar \gamma }_{{i_{{N_c,1}}}}}}}\!\sum\limits_{\substack{
 {i_{{N_c} + 1,1}}, \cdots ,{i_{N_1,1}} \\
 {i_{{N_c} + 1,1}} \ne  \cdots  \ne {i_{N_1,1}} \\
  \vdots  \\
 {i_{N_1,1}} \ne {i_{{N_c} + 1,1}} \\
 }}^{1,2, \cdots {N_1}} \!{ \sum\limits_{\left\{\! {{i_{1,1}}, \cdots ,{i_{{N_c} - 1,1}}} \!\right\} \in {P_{{N_c} - 1}}\!\left( \!{{I_{N_1}} - \left\{\! {{i_{{N_c,1}}}} \!\right\} - \left\{\! {{i_{{N_c} + 1,1}}, \ldots ,{i_{N_1,1}}} \!\right\}} \!\right)} {\prod\limits_{\scriptstyle q = 1 \atop
  \scriptstyle \left\{\! {{i_{1,1}}, \cdots ,{i_{{N_c} - 1,1}}} \!\right\}}^{{N_c}} \!\!\!\!{{C_{q,1,{N_c} - 1}}} } } }
\\
&\times \left[ {\frac{{ - 1}}{{\left( {\sum\limits_{l = 1}^{{N_c}} {\frac{1}{{{{\bar \gamma }_{{i_{l,1}}}}}} - \frac{{{N_c}}}{{{{\bar \gamma }_{{i_{q,1}}}}}}} } \right)}}\left\{ {{{\bar \gamma }_{{i_{q,1}}}}\left( {1 - \exp \left( { - \frac{x}{{{{\bar \gamma }_{{i_{q,1}}}}}}} \right)} \right) - \frac{{{N_c}}}{{\sum\limits_{l = 1}^{{N_c}} {\frac{1}{{{{\bar \gamma }_{{i_{l,1}}}}}}} }}\left( {1 - \exp \left( { - \left( {\sum\limits_{l = 1}^{{N_c}} {\frac{1}{{{{\bar \gamma }_{{i_{l,1}}}}}}} } \right)\frac{x}{{{N_c}}}} \right)} \right)} \right\}} \right.
\\
&\quad\quad- \prod\limits_{\scriptstyle k' = 1 \atop
  \scriptstyle \left\{ {{i_{{N_c} + 1,1}}, \cdots ,{i_{N_1,1}}} \right\}}^{{N_1} - {N_c}} {{{\left( { - 1} \right)}^{k'}}\sum\limits_{{j_1} = {j_0} + {N_c} + 1}^{{N_1} - k' + 1} { \cdots \sum\limits_{{j_{k'}} = {j_{k' - 1}} + 1}^{N_1} {\frac{1}{{\left( {\sum\limits_{l = 1}^{{N_c}} {\frac{1}{{{{\bar \gamma }_{{i_{l,1}}}}}} + \sum\limits_{m = 1}^{k'} {\frac{1}{{{{\bar \gamma }_{{i_{{j_m,1}}}}}}} - \frac{{{N_c}}}{{{{\bar \gamma }_{{i_{q,1}}}}}}} } } \right)}}} } }
\\
&\quad\quad\left. { \times \left\{ {{{\bar \gamma }_{{i_{q,1}}}}\left( {1 - \exp \left( { - \frac{x}{{{{\bar \gamma }_{{i_{q,1}}}}}}} \right)} \right) - \frac{{{N_c}}}{{\sum\limits_{l = 1}^{{N_c}} {\frac{1}{{{{\bar \gamma }_{{i_{l,1}}}}}} + \sum\limits_{m = 1}^{k'} {\frac{1}{{{{\bar \gamma }_{{i_{{j_m,1}}}}}}}} } }}\left( {1 - \exp \left( { - \left( {\sum\limits_{l = 1}^{{N_c}} {\frac{1}{{{{\bar \gamma }_{{i_{l,1}}}}}} + \sum\limits_{m = 1}^{k'} {\frac{1}{{{{\bar \gamma }_{{i_{{j_m,1}}}}}}}} } } \right)\frac{x}{{{N_c}}}} \right)} \right)} \right\}} \right],
\end{aligned}
\end{equation}
where $j_0=0$,
\begin{equation} \scriptsize  \label{eq:15}
{C_{l,n_1,n_2}} = \frac{1}{{\prod\limits_{l = {n_1}}^{{n_2}} {\left( { - {{\bar \gamma }_{{i_{l,1}}}}} \right)} }{F'_{l,n_1,n_2}\left( {\frac{1}{{{{\bar \gamma }_{{i_{l,1}}}}}}} \right)}},
\end{equation}
\begin{equation} \scriptsize  \label{eq:16}
\!\!\!\!F'_{l,n_1,n_2}\left( x \right) \!= \!\Bigg[\! {\sum\limits_{l = 1}^{{n_2} - {n_1}}\! {\left(\! {{n_2} - {n_1} - l + 1} \!\right){x^{{n_2} - {n_1} - l}}{{\left(\! { - 1} \!\right)}^l}}}{{\sum\limits_{{j_1} = {j_0} + {n_1}}^{{n_2} - l + 1} \!\!{ \cdots \!\!\sum\limits_{{j_l} = {j_{l - 1}} + 1}^{{n_2}} {\prod\limits_{m = 1}^l {\frac{1}{{{{\bar \gamma }_{{i_{{j_m,1}}}}}}}} } } } } \!\Bigg] \!+ \!\left(\! {{n_2} - {n_1} + 1} \!\right){x^{{n_2} - {n_1}}}.
\end{equation}

\begin{landscape}
\subsection{Joint PDF of Two Adjacent Partial Sums $Y$ and $W_1$ over i.n.d. Rayleigh Fading, ${{f_{Y,{W_1}}}\left( x, y \right)}$}
\begin{equation} \label{eq:17}
\scriptsize
\!\!\!\!\!\!\begin{aligned}
&{f_{Y,{W_1}}}\!\left( \!{x,y}\! \right)\! = \!\sum\limits_{\scriptstyle{i_{{N_c},1}}, \ldots ,{i_{{N_1},1}}\atop
\scriptstyle{i_{{N_c},1}} \ne  \cdots  \ne {i_{{N_1},1}}}^{1,2, \ldots ,{N_1}} \!\!{\frac{1}{{{{\bar \gamma }_{{i_{{N_c},1}}}}}}\!\sum\limits_{\scriptstyle{i_{{N_c} - {N_s},1}} = 1\atop
\scriptstyle{i_{{N_c} - {N_s},1}} \ne {i_{{N_c},1, \ldots ,}}{i_{{N_1},1}}}^{{N_1}}\!\!\! {\frac{1}{{{{\bar \gamma }_{{i_{{N_c} - {N_s},1}}}}}}\!\sum\limits_{\left\{ {{i_{{N_c} - {N_s} + 1,1}}, \ldots ,{i_{{N_c} - 1,1}}} \right\} \in {{\mathop{\rm P}\nolimits} _{{N_s} - 1}}\left(\! {{I_{{N_1}}} \!- \!\left\{ {{i_{{N_c} - {N_s},1}}} \!\right\}\! -\! \left\{ \!{{i_{{N_c},1}}, \ldots ,{i_{{N_1},1}}} \!\right\}} \!\right)}\! {\sum\limits_{k = {N_c}\! -\! {N_s}\! +\! 1}^{{N_c}\! - \!1} \!{{C_{k,{N_c} - {N_s} + 1,{N_c} - 1}}} } } }
\\
&\left[ \!{\sum\limits_{\left\{ {{i_{1,1}}, \ldots ,{i_{{N_c} - {N_s} - 1,1}}} \right\} \in {{\mathop{\rm P}\nolimits} _{{N_c} - {N_s} - 1}}\left( {{I_{{N_1}}} - \left\{ {{i_{{N_c} - {N_s},1}}} \right\} - \left\{ {{i_{{N_c},1}}, \ldots ,{i_{{N_1},1}}} \right\} - \left\{ {{i_{{N_c} - {N_s} + 1,1}}, \ldots ,{i_{{N_c} - 1,1}}} \right\}} \right)}\! {\sum\limits_{h = 1}^{{N_c} \!- \!{N_s}\! - \!1}\! {{C_{h,1,{N_c} - {N_s} - 1}}} } } \right.\!\exp \left(\! { -\! \frac{x}{{{{\bar \gamma }_{{i_{h,1}}}}}}} \!\right)\!\exp \left( \!{ - \!\frac{y}{{{{\bar \gamma }_{{i_{k,1}}}}}}} \!\right)
\\
& \times {{\bf{{\rm I}}}}\left( {{z_2},\beta ,\frac{y}{{{N_s}}},\frac{x}{{{N_c} - {N_s}}};{z_4},\alpha ,0,\frac{y}{{{N_s}}}} \right)
\\
& +\! \sum\limits_{l = 1}^{{N_s}\! - \!1}\! {{{\left(\! { - 1}\! \right)}^l}\!\sum\limits_{{j_1} = {j_0} \!+ \!{N_c}\! - \!{N_s}\! + \!1}^{{N_c} \!- \!l} { \cdots \sum\limits_{{j_l} = {j_{l - 1}}\! + \!1}^{{N_c} \!- \!1}\! {\sum\limits_{\left\{ {{i_{1,1}}, \ldots ,{i_{{N_c} - {N_s} - 1,1}}} \right\} \in {{\mathop{\rm P}\nolimits} _{{N_c} - {N_s} - 1}}\left( {{I_{{N_1}}}\! - \!\left\{ {{i_{{N_c} - {N_s},1}}} \right\}\! - \!\left\{ {{i_{{N_c},1}}, \ldots ,{i_{{N_1},1}}} \right\} \!- \!\left\{ {{i_{{N_c} - {N_s} + 1,1}}, \ldots ,{i_{{N_c} - 1,1}}} \right\}} \right)}\! {\sum\limits_{h = 1}^{{N_c}\!- \!{N_s}\! - \!1} {{C_{h,1,{N_c} - {N_s} - 1}}} } } } } \!\exp \!\left(\! { - \!\frac{x}{{{{\bar \gamma }_{{i_h}}}}}} \!\right)\!\exp\! \left(\! { - \!\frac{y}{{{{\bar \gamma }_{{i_k}}}}}}\! \right)
\\
&\left. { \times \!\left\{\! {{\bf{{\rm I}}}\!\left(\! {{z_2},\beta ',\frac{y}{{{N_s}}},\frac{x}{{{N_c}\! - \!{N_s}}};{z_4},\alpha '',0,\min\! \left[\! {\frac{y}{{{N_s}}},\frac{{y\! -\! \frac{l}{{{N_c} \!-\! {N_s}}} \cdot x}}{{\left(\! {{N_s} \!-\! l} \!\right)}}}\! \right]} \!\right) \!+\! {\bf{{\rm I}'}}\!\left(\! {{z_2},\beta ',\frac{y}{{{N_s}}},\frac{{y \!- \!\left(\! {{N_s} \!- \!l} \!\right) \cdot {z_4}}}{l};{z_4},\alpha '',0,\frac{y}{{{N_s}}}} \right)\left. { - \!{\bf{{\rm I}'}}\!\left(\! {{z_2},\beta ',\frac{y}{{{N_s}}},\frac{{y \!- \!\left(\! {{N_s}\! - \!l} \!\right) \cdot {z_4}}}{l};{z_4},\alpha '',0,\min \!\left[\! {\frac{y}{{{N_s}}},\frac{{y\! - \!\frac{l}{{{N_c} \!- \!{N_s}}} \cdot x}}{{\left(\! {{N_s}\! - \!l} \!\right)}}} \!\right]} \!\right)} \!\right\}} \right.}\! \right]
\\
&+ \sum\limits_{\scriptstyle{i_{{N_c},1}}, \ldots ,{i_{{N_1},1}}\atop
\scriptstyle{i_{{N_c},1}} \ne  \cdots  \ne {i_{{N_1},1}}}^{1,2, \ldots ,{N_1}} {\frac{1}{{{{\bar \gamma }_{{i_{{N_c},1}}}}}}\sum\limits_{g = 1}^{{N_1} - {N_c}} {{{\left( { - 1} \right)}^g}\sum\limits_{{{j'}_1} = {{j'}_0} + {N_c} + 1}^{{N_1} - g + 1} { \cdots \sum\limits_{{{j'}_g} = {{j'}_{g - 1}} + 1}^{{N_1}} {\sum\limits_{\scriptstyle{i_{{N_c} - {N_s},1}} = 1\atop
\scriptstyle{i_{{N_c} - {N_s},1}} \ne {i_{{N_c},1, \ldots ,}}{i_{{N_1},1}}}^{{N_1}} {\frac{1}{{{{\bar \gamma }_{{i_{{N_c} - {N_s},1}}}}}}} } } } } \sum\limits_{\left\{ {{i_{{N_c} - {N_s} + 1,1}}, \ldots ,{i_{{N_c} - 1,1}}} \right\} \in {{\mathop{\rm P}\nolimits} _{{N_s} - 1}}\left( {{I_{{N_1}}} - \left\{ {{i_{{N_c} - {N_s},1}}} \right\} - \left\{ {{i_{{N_c},1}}, \ldots ,{i_{{N_1},1}}} \right\}} \right)} {}
\\
&\sum\limits_{k = {N_c} - {N_s} + 1}^{{N_c} - 1} {{C_{k,{N_c} - {N_s} + 1,{N_c} - 1}}}
\left[ {\sum\limits_{\left\{ {{i_{1,1}}, \ldots ,{i_{{N_c} - {N_s} - 1,1}}} \right\} \in {{\mathop{\rm P}\nolimits} _{{N_c} - {N_s} - 1}}\left( {{I_{{N_1}}} - \left\{ {{i_{{N_c} - {N_s},1}}} \right\} - \left\{ {{i_{{N_c},1}}, \ldots ,{i_{{N_1},1}}} \right\} - \left\{ {{i_{{N_c} - {N_s} + 1,1}}, \ldots ,{i_{{N_c} - 1,1}}} \right\}} \right)} {\sum\limits_{h = 1}^{{N_c} - {N_s} - 1} {{C_{h,1,{N_c} - {N_s} - 1}}\!\exp \!\left(\! { - \!\frac{x}{{{{\bar \gamma }_{{i_h}}}}}}\! \right)\!\exp \!\left(\! { -\! \frac{y}{{{{\bar \gamma }_{{i_k}}}}}} \!\right)} } } \right.
\\
&\times {{\bf{{\rm I}}}}\left( {{z_2},\beta ,\frac{y}{{{N_s}}},\frac{x}{{{N_c} - {N_s}}};{z_4},\alpha ',0,\frac{y}{{{N_s}}}} \right)
\\
&+\! \sum\limits_{l = 1}^{{N_s}\! - \!1} \!{{{\left(\! { - 1} \!\right)}^l}\!\sum\limits_{{j_1} = {j_0} \!+ \!{N_c}\! -\! {N_s} \!+ 1}^{{N_c}\! -\! l} { \cdots \sum\limits_{{j_l} = {j_{l - 1}}\! + \!1}^{{N_c}\! - \!1} \!{\sum\limits_{\left\{ {{i_{1,1}}, \ldots ,{i_{{N_c} - {N_s} - 1,1}}} \right\} \in {{\mathop{\rm P}\nolimits} _{{N_c} - {N_s} - 1}}\left( {{I_{{N_1}}} \!- \!\left\{ {{i_{{N_c} - {N_s},1}}} \right\} \!- \!\left\{ {{i_{{N_c},1}}, \ldots ,{i_{{N_1},1}}} \right\} \!-\! \left\{ {{i_{{N_c} - {N_s} + 1,1}}, \ldots ,{i_{{N_c} - 1,1}}} \right\}} \right)}\! {\sum\limits_{h = 1}^{{N_c} \!- \!{N_s}\! - \!1} {{C_{h,1,{N_c} - {N_s} - 1}}\!\exp \!\left(\! { -\! \frac{x}{{{{\bar \gamma }_{{i_h}}}}}} \!\right)\!\exp \!\left(\! { - \!\frac{y}{{{{\bar \gamma }_{{i_k}}}}}} \!\right)} } } } }
\\
&\left. { \times \!\left\{ \!{{\bf{{\rm I}}}\!\left(\! {{z_2},\beta ',\frac{y}{{{N_s}}},\frac{x}{{{N_c}\! - \!{N_s}}};{z_4},\alpha ''',0,\min \!\left[ \!{\frac{y}{{{N_s}}},\frac{{y \!-\! \frac{l}{{{N_c}\! -\! {N_s}}} \cdot x}}{{\left( {{N_s}\! - \!l} \right)}}}\! \right]} \!\right)\! +\! {\bf{{\rm I}'}}\!\left(\! {{z_2},\beta ',\frac{y}{{{N_s}}},\frac{{y \!- \!\left(\! {{N_s} \!- \!l} \!\right) \cdot {z_4}}}{l};{z_4},\alpha ''',0,\frac{y}{{{N_s}}}} \right)\left. { - \!{\bf{{\rm I}'}}\!\left(\! {{z_2},\beta ',\frac{y}{{{N_s}}},\frac{{y \!- \!\left(\! {{N_s}\! -\! l} \!\right) \cdot {z_4}}}{l};{z_4},\alpha ''',0,\min \!\left[\! {\frac{y}{{{N_s}}},\frac{{y \!-\! \frac{l}{{{N_c}\! - \!{N_s}}} \cdot x}}{{\left(\! {{N_s}\! -\! l}\! \right)}}}\! \right]} \!\right)} \!\right\}} \right.} \!\right],
\end{aligned}
\end{equation}
\end{landscape}
where
\small$\alpha  =  - \left( {\sum\limits_{l = {N_c-N_s} + 1}^{{N_c}} {\left( {\frac{1}{{{{\bar \gamma }_{{i_{l,1}}}}}}} \right) - \frac{{\left( {{N_s}} \right)}}{{{{\bar \gamma }_{{i_{k,1}}}}}}} } \right)$\normalsize, \small$\alpha ' =  - \Bigg( \sum\limits_{l = {N_c-N_s} + 1}^{{N_c}} \left( {\frac{1}{{{{\bar \gamma }_{{i_{l,1}}}}}}} \right) + \sum\limits_{m = 1}^g {\frac{1}{{{{\bar \gamma }_{{i_{{{j'}_m,1}}}}}}}}  - \frac{{\left( {{N_s}} \right)}}{{{{\bar \gamma }_{{i_{k,1}}}}}}  \Bigg)$\normalsize, \small$\alpha '' =  - \left( {\sum\limits_{l = {N_c-N_s} + 1}^{{N_c}} {\left( {\frac{1}{{{{\bar \gamma }_{{i_{{l,1}}}}}}}} \right) - \frac{{\left( {{N_s} - l} \right)}}{{{{\bar \gamma }_{{i_{{k,1}}}}}}} - \sum\limits_{q = 1}^l {\frac{1}{{{{\bar \gamma }_{{i_{{j_q,1}}}}}}}} } } \right)$\normalsize, \small$\alpha ''' =  - \Bigg( \sum\limits_{l = {N_c-N_s} + 1}^{N_c} \left( \frac{1}{{\bar \gamma }_{i_{{l,1}}}} \right) + \sum\limits_{m = 1}^g {\frac{1}{{{{\bar \gamma }_{{i_{{{j'}_m,1}}}}}}}}  - \frac{{\left( {{N_s} - l} \right)}}{{{{\bar \gamma }_{{i_{{k,1}}}}}}} - \sum\limits_{q = 1}^l {\frac{1}{{{{\bar \gamma }_{{i_{{j_q,1}}}}}}}}   \Bigg)$\normalsize, \small$\beta  =  - \Bigg( \sum\limits_{l = 1}^{{N_c-N_s}} \Big( \frac{1}{{{{\bar \gamma }_{i_{l,1}}}}} \Big) - \frac{{N_c-N_s}}{{{{\bar \gamma }_{{i_{h,1}}}}}}  \Bigg)$\normalsize, and \small$\beta'  =  - \left( {\sum\limits_{l = 1}^{N_c-N_s} {\left( {\frac{1}{{{{\bar \gamma }_{{i_{{l,1}}}}}}}} \right) + \sum\limits_{q = 1}^l {\frac{1}{{{{\bar \gamma }_{{i_{{j_q,1}}}}}}}}  - \frac{{N_c-N_s}}{{{{\bar \gamma }_{{i_{{h,1}}}}}}} - \frac{l}{{{{\bar \gamma }_{{i_{k,1}}}}}}} } \right)$

\subsection{CDF of the Sums of the $N_s$ Strongest Paths from Each Target BS over i.n.d. Rayleigh Fading, ${{F_{{W_n}}}\left( x \right)}$}
\begin{equation} \scriptsize \label{eq:19}
\begin{aligned}
F_{W_n}\left( x \right) \!=& \!\sum\limits_{{i_{{N_s,n}}} = 1}^{N_n}\! {\frac{1}{{{{\bar \gamma }_{{i_{{N_s,n}}}}}}}\!\sum\limits_{\substack{
 {i_{{N_s} + 1,n}}, \cdots ,{i_{N_n,n}} \\
 {i_{{N_s} + 1,n}} \ne  \cdots  \ne {i_{N_n,n}} \\
  \vdots  \\
 {i_{N_n,n}} \ne {i_{{N_s} + 1,n}} \\
 }}^{1,2, \cdots {N_n}}\! {\sum\limits_{\left\{\! {{i_{1,n}}, \cdots ,{i_{{N_s} - 1,n}}}\! \right\} \in {P_{{N_s} - 1}}\left(\! {{I_{N_n}} - \left\{\! {{i_{{N_s,n}}}} \!\right\} - \left\{ \!{{i_{{N_s} + 1,n}}, \ldots ,{i_{N_n,n}}}\! \right\}} \!\right)} {\prod\limits_{\scriptstyle q = 1 \atop
  \scriptstyle \left\{\! {{i_{1,n}}, \cdots ,{i_{{N_s} - 1,n}}}\! \right\}}^{{N_s}}\!\!\!\!\! {{C_{q,1,{N_s} - 1}}} } } }
\\
&\times \left[ {\frac{{ - 1}}{{\left( {\sum\limits_{l = 1}^{{N_s}} {\frac{1}{{{{\bar \gamma }_{{i_{l,n}}}}}} - \frac{{{N_s}}}{{{{\bar \gamma }_{{i_{q,n}}}}}}} } \right)}}\left\{ {{{\bar \gamma }_{{i_{q,n}}}}\left( {1 - \exp \left( { - \frac{x}{{{{\bar \gamma }_{{i_{q,n}}}}}}} \right)} \right) - \frac{{{N_s}}}{{\sum\limits_{l = 1}^{{N_s}} {\frac{1}{{{{\bar \gamma }_{{i_{l,n}}}}}}} }}\left( {1 - \exp \left( { - \left( {\sum\limits_{l = 1}^{{N_s}} {\frac{1}{{{{\bar \gamma }_{{i_{l,n}}}}}}} } \right)\frac{x}{{{N_s}}}} \right)} \right)} \right\}} \right.
\\
&\quad\quad- \prod\limits_{\scriptstyle k' = 1 \atop
  \scriptstyle \left\{ {{i_{{N_s} + 1,n}}, \cdots ,{i_{N_n,n}}} \right\}}^{{N_n} - {N_s}} {{{\left( { - 1} \right)}^{k'}}\sum\limits_{{j_1} = {j_0} + {N_s} + 1}^{{N_n} - k' + 1} { \cdots \sum\limits_{{j_{k'}} = {j_{k' - 1}} + 1}^{N_n} {\frac{1}{{\left( {\sum\limits_{l = 1}^{{N_s}} {\frac{1}{{{{\bar \gamma }_{{i_{l,n}}}}}} + \sum\limits_{m = 1}^{k'} {\frac{1}{{{{\bar \gamma }_{{i_{{j_m,n}}}}}}} - \frac{{{N_s}}}{{{{\bar \gamma }_{{i_{q,n}}}}}}} } } \right)}}} } }
\\
& \quad\quad\quad\left. \sizecorr{\times \left[\! {\frac{{ - 1}}{{\left(\! {\sum\limits_{l = 1}^{{N_s}}\! {\frac{1}{{{{\bar \gamma }_{{i_{l,n}}}}}}\! - \!\frac{{{N_s}}}{{{{\bar \gamma }_{{i_{q,n}}}}}}} }\! \right)}}\!\left\{\! {{{\bar \gamma }_{{i_{q,n}}}}\!\left(\! {1 - \exp \!\left( \!{ - \frac{x}{{{{\bar \gamma }_{{i_{q,n}}}}}}} \!\right)} \!\right) - \frac{{{N_s}}}{{\sum\limits_{l = 1}^{{N_s}} \!{\frac{1}{{{{\bar \gamma }_{{i_{l,n}}}}}}} }}\!\left( \!{1 - \exp \!\left(\! { - \left(\! {\sum\limits_{l = 1}^{{N_s}}\! {\frac{1}{{{{\bar \gamma }_{{i_{l,n}}}}}}} }\! \right)\frac{x}{{{N_s}}}} \!\right)} \!\right)}\! \right\}} \right.}
{ \times \left\{\! {{{\bar \gamma }_{{i_{q,n}}}}\left(\! {1 - \exp \!\left(\! { - \frac{x}{{{{\bar \gamma }_{{i_{q,n}}}}}}} \!\right)}\! \right) \!- \!\frac{{{N_s}}}{{\sum\limits_{l = 1}^{{N_s}} \!{\frac{1}{{{{\bar \gamma }_{{i_{l,n}}}}}} \!+ \!\sum\limits_{m = 1}^{k'} \!{\frac{1}{{{{\bar \gamma }_{{i_{{j_m,n}}}}}}}} } }}\!\left(\! {1 - \exp\! \left(\! { - \left(\! {\sum\limits_{l = 1}^{{N_s}}\! {\frac{1}{{{{\bar \gamma }_{{i_{l,n}}}}}} \!+ \!\sum\limits_{m = 1}^{k'} \!{\frac{1}{{{{\bar \gamma }_{{i_{{j_m,n}}}}}}}} } } \!\right)\frac{x}{{{N_s}}}} \!\right)} \!\right)} \!\right\}} \!\right].
\end{aligned}
\end{equation}\normalsize

Note that in this report, we provide all three required key
statistics in (\ref{eq:14}), (\ref{eq:17}), and  (\ref{eq:19}), in
the closed-form expressions to accurately investigating the
performance measures mentioned in Sec.~\ref{sec_2}, especially, over
i.n.d. Rayleigh fading conditions while \cite{kn:S_Choi_2008_1}
provides non-closed-form expressions even over i.i.d. fading
assumptions since the final results involve finite integrations. With these joint
statistics derived in closed-form expressions, the outage
probability as well as other performance measures mentioned in
Sec.~\ref{sec_2} can be easily calculated with standard mathematical
softwares such as Mathematica.

\section{Conclusions}\label{conc}
In this work, we studied the assessment tool of the finger
replacement scheme proposed in~\cite{kn:S_Choi_2008_1} over i.n.d.
fading conditions by providing the general comprehensive
mathematical framework with non-identical parameters. Specifically,
we provided the closed-form expressions for the required key
statistics of i.n.d. ordered exponential RVs by applying a unified
framework to determine the target statistics of partial sums of
ordered RVs proposed in \cite{kn:sungsiknam2013_ISIT,
kn:IND_MGF_sungsiknam_1} and the general comprehensive framework for
the outage performance based on the these statistical results. The
proposed approach is quite general to apply to the performance
analysis of various wireless communication systems over practical
fading channels.

In Fig. 2, we assess the effect of non-identically distributed paths
on the outage performance of the replacement schemes. More
specifically, instead of the uniform power delay profile (PDP)
considered so far, we now consider an exponentially decaying PDP.
More specifically, we assume that the channel has an exponential
multipath intensity profile (MIP), for which
$\bar\gamma_{i}=\bar\gamma\cdot\exp\left(-\delta\left(i-1\right)\right)$,
$\left( 1\le i \le N_n, 1\le n \le L\right)$ where $\bar\gamma_{i}$
is the average SNR of the $i$-th path out of the total available
resolvable paths from each BS, $\bar\gamma$ is the strongest average
SNR (or the average SNR of the first path), and $\delta$ is the
power decay factor. Note that $\delta=0$ means identically
distributed paths. These results show that the effect of path
unbalance induces non-negligible performance degradation compared
with the results for i.i.d. fading scenario. \cite{kn:S_Choi_2008_1}
showed that the proposed scheme can be still applied to the i.n.d.
fading scenario. However, this effect must be taken into account for
the accurate prediction of the performance over i.n.d. fading
environments and with our analytical results, we believe that it is
available to accurately predict the performance.

\newpage
{\section*{Appendices}
\appendices
In here, for analytical convenience, we assume that $N_n=N$, $u_{i,n}=u_i$, and ${\bar \gamma }_{i_{l,n}}={\bar \gamma }_{i_{l}}$ for all $n=1,2,\cdots,L$.

\section{Derivation of (\ref{eq:7})} \label{appendix_1}
Based on the mode of operation, we need to consider two cases i) the final combined SNR is greater than or equal to the target SNR, $\gamma_T$ and ii) the
final combined SNR falls below $\gamma_T$, separately. For case ii), after scanning the paths from the serving BS as well as all the target BSs, the combined SNR of the resolvable paths from the serving BS and all the target BS falls below $\gamma_T$. Therefore, we can directly re-write (\ref{eq:3}) for $0 < x < \gamma_T$ as
\begin{equation} \label{APP:1}
\begin{aligned}
{F_{{\gamma _F}}}\left( x \right) =\Pr \left[ {Y + \max \left\{ {{W_1},{W_2}, \cdots ,{W_L}} \right\} < x} \right].
\end{aligned}
\end{equation}
However, for case i), we also need to consider two cases separately a) $Y + {W_1} \ge \gamma_T$ and b) $Y + {W_1} < \gamma_T$. More specifically, for case a), no finger replacement is needed while for case b), the receiver attempts a two-way SHO by starting to scan additional paths from the target BSs and the final combined SNR should be greater than or equal to $\gamma_T$. By considering the case ii)-a) and ii)-b), we can write (\ref{eq:3}) for $x \ge \gamma_T$ as
\begin{equation} \label{APP:2}
\begin{aligned}
{F_{{\gamma _F}}}\left( x \right) =&  \Pr \left[ Y + {W_1}\ge{\gamma _T}, Y + {W_1} < x \right]
\\
&+ \Pr \left[ {Y + {W_1} < {\gamma _T},{\gamma _T} \le Y + \max \left\{ {{W_1},{W_2}, \cdots ,{W_L}} \right\} < x} \right].
\end{aligned}
\end{equation}
As results, after some manipulations, we can re-write (\ref{APP:2}) in the simplified form given in (\ref{eq:7}) for $x \ge {\gamma _T}$.

\section{CDF of the ${N_c}/{N}$-GSC output SNR} \label{appendix_2}
Based on the proposed unified frame work in  \cite{kn:unified_approach}, noting that  $Z'=Y+W_1$, we can obtain the target CDF of $Z'=\sum\limits_{i=1}^{N_c} {u_{i}}$ with the 2-dimensional joint PDF of $Z_1=\sum\limits_{i=1}^{N_c-1} {u_{i}}$ and $Z_2={u_{N_c}}$. Specifically, by letting $X=Z_1+Z_2$, we can obtain the target CDF of $Z^{'}=X$ by integrating over $z_2$ for a given condition $Z_2\le \frac{X}{N_c}$ yielding (\ref{eq:11}). Fortunately, by adopting \cite[Eq. (51)]{kn:IND_MGF_sungsiknam_1} to (\ref{eq:11}), we can obtain the closed-form expression of (\ref{eq:11}) over i.n.d. Rayleigh fading conditions by performing the double integrations over $z_2$ and $z$ in order.

After inserting \cite[Eq. (51)]{kn:IND_MGF_sungsiknam_1} in (\ref{eq:11}), the inner integral term in (\ref{eq:11}) can be re-written as
\begin{equation} \footnotesize \label{eq:20}
\begin{aligned}
&\int_0^{\frac{z}{{{N_c}}}} {{f_Z}\left( {z - {z_2},{z_2}} \right)d{z_2}}  = \sum\limits_{{i_{{N_c}}} = 1}^N \!{\frac{1}{{{{\bar \gamma }_{{i_{{N_c}}}}}}}\!\sum\limits_{\substack{
 {i_{{N_c} + 1}}, \cdots ,{i_N} \\
 {i_{{N_c} + 1}} \ne  \cdots  \ne {i_N} \\
  \vdots  \\
 {i_N} \ne {i_{{N_c} + 1}} \\
 }}^{1,2, \cdots N}\! {\sum\limits_{\left\{\! {{i_1}, \cdots ,{i_{{N_c} - 1}}} \!\right\} \in {P_{{N_c} - 1}}\left( \!{{I_N} - \left\{\! {{i_{{N_c}}}} \!\right\} - \left\{\! {{i_{{N_c} + 1}}, \ldots ,{i_N}} \!\right\}} \!\right)} {\prod\limits_{\scriptstyle q = 1 \atop
  \scriptstyle \left\{\! {{i_1}, \cdots ,{i_{{N_c} - 1}}} \!\right\}}^{{N_c}}\!\!\! {{C_{q,1,{N_c} - 1}}} } } }
\\
&\times \left[ { - \int_0^{\frac{z}{{{N_c}}}} {\exp \left( { - \frac{z}{{{{\bar \gamma }_{{i_q}}}}} - \left( {\sum\limits_{l = 1}^{{N_c}} {\frac{1}{{{{\bar \gamma }_{{i_l}}}}} - \frac{{{N_c}}}{{{{\bar \gamma }_{{i_q}}}}}} } \right){z_2}} \right)d{z_2}} } \sizecorr{\left. { - \prod\limits_{\scriptstyle k' = 1 \atop
  \scriptstyle \left\{ {{i_{{N_c} + 1}}, \cdots ,{i_N}} \right\}}^{N - {N_c}} {{{\left( { - 1} \right)}^{k'}}\sum\limits_{{j_1} = {j_0} + {N_c} + 1}^{N - k' + 1} { \cdots \sum\limits_{{j_{k'}} = {j_{k' - 1}} + 1}^N {\int_0^{\frac{z}{{{N_c}}}} {\exp \left( { - \frac{z}{{{{\bar \gamma }_{{i_q}}}}} - \left( {\sum\limits_{l = 1}^{{N_c}} {\frac{1}{{{{\bar \gamma }_{{i_l}}}}} + \sum\limits_{m = 1}^{k'} {\frac{1}{{{{\bar \gamma }_{{i_{{j_m}}}}}}} - \frac{{{N_c}}}{{{{\bar \gamma }_{{i_q}}}}}} } } \right){z_2}} \right)d{z_2}} } } } } \right].}\right.
\\
&\quad\quad\left. { - \prod\limits_{\scriptstyle k' = 1 \atop
  \scriptstyle \left\{ {{i_{{N_c} + 1}}, \cdots ,{i_N}} \right\}}^{N - {N_c}} {{{\left( { - 1} \right)}^{k'}}\sum\limits_{{j_1} = {j_0} + {N_c} + 1}^{N - k' + 1} { \cdots \sum\limits_{{j_{k'}} = {j_{k' - 1}} + 1}^N {\int_0^{\frac{z}{{{N_c}}}} {\exp \left( { - \frac{z}{{{{\bar \gamma }_{{i_q}}}}} - \left( {\sum\limits_{l = 1}^{{N_c}} {\frac{1}{{{{\bar \gamma }_{{i_l}}}}} + \sum\limits_{m = 1}^{k'} {\frac{1}{{{{\bar \gamma }_{{i_{{j_m}}}}}}} - \frac{{{N_c}}}{{{{\bar \gamma }_{{i_q}}}}}} } } \right){z_2}} \right)d{z_2}} } } } } \right].
\end{aligned}
\end{equation}
In (\ref{eq:20}), the first and second integral terms can be evaluated as the following closed-form expressions with the help of  the basic exponential integration \cite{kn:abramowitz}
\begin{equation} \small \label{eq:21}
\frac{1}{{\left( {\sum\limits_{l = 1}^{{N_c}} {\frac{1}{{{{\bar \gamma }_{{i_l}}}}} - \frac{{{N_c}}}{{{{\bar \gamma }_{{i_q}}}}}} } \right)}}\left[ {\exp \left( { - \frac{z}{{{{\bar \gamma }_{{i_q}}}}}} \right) - \exp \left( { - \left( {\sum\limits_{l = 1}^{{N_c}} {\frac{1}{{{{\bar \gamma }_{{i_l}}}}}} } \right)\frac{z}{{{N_c}}}} \right)} \right],
\end{equation}
\begin{equation} \small \label{eq:22}
\frac{1}{{\left( {\sum\limits_{l = 1}^{{N_c}} {\frac{1}{{{{\bar \gamma }_{{i_l}}}}} + \sum\limits_{m = 1}^{k'} {\frac{1}{{{{\bar \gamma }_{{i_{{j_m}}}}}}} - \frac{{{N_c}}}{{{{\bar \gamma }_{{i_q}}}}}} } } \right)}}\left[ {\exp \left( { - \frac{z}{{{{\bar \gamma }_{{i_q}}}}}} \right) - \exp \left( { - \left( {\sum\limits_{l = 1}^{{N_c}} {\frac{1}{{{{\bar \gamma }_{{i_l}}}}} + \sum\limits_{m = 1}^{k'} {\frac{1}{{{{\bar \gamma }_{{i_{{j_m}}}}}}}} } } \right)\frac{z}{{{N_c}}}} \right)} \right].
\end{equation}

Subsequently, after substituting (\ref{eq:21}) and (\ref{eq:22}) in (\ref{eq:20}), we can obtain a closed-form expression of the inner integral term in (\ref{eq:11}) as
\begin{equation} \footnotesize \label{eq:23}
\!\!\!\!\!\!\!\!\!\!\!\!\!\!\begin{aligned}
&{f_{Z'}}\left( z \right) = \sum\limits_{{i_{{N_c}}} = 1}^N {\frac{1}{{{{\bar \gamma }_{{i_{{N_c}}}}}}}\sum\limits_{\substack{
 {i_{{N_c} + 1}}, \cdots ,{i_N} \\
 {i_{{N_c} + 1}} \ne  \cdots  \ne {i_N} \\
  \vdots  \\
 {i_N} \ne {i_{{N_c} + 1}} \\
 }}^{1,2, \cdots N} { \sum\limits_{\left\{ {{i_1}, \cdots ,{i_{{N_c} - 1}}} \right\} \in {P_{{N_c} - 1}}\left( {{I_N} - \left\{ {{i_{{N_c}}}} \right\} - \left\{ {{i_{{N_c} + 1}}, \ldots ,{i_N}} \right\}} \right)} {\prod\limits_{\scriptstyle q = 1 \atop
  \scriptstyle \left\{ {{i_1}, \cdots ,{i_{{N_c} - 1}}} \right\}}^{{N_c}} {{C_{q,1,{N_c} - 1}}} } } }
\\
&\times \left[ {\frac{{ - 1}}{{\left( {\sum\limits_{l = 1}^{{N_c}} {\frac{1}{{{{\bar \gamma }_{{i_l}}}}} - \frac{{{N_c}}}{{{{\bar \gamma }_{{i_q}}}}}} } \right)}}\left\{ {\exp \left( { - \frac{z}{{{{\bar \gamma }_{{i_q}}}}}} \right) - \exp \left( { - \left( {\sum\limits_{l = 1}^{{N_c}} {\frac{1}{{{{\bar \gamma }_{{i_l}}}}}} } \right)\frac{z}{{{N_c}}}} \right)} \right\}} \right.
\\
&
\left. { - \!\!\!\!\!\!\prod\limits_{\scriptstyle k' = 1 \atop
  \scriptstyle \left\{\! {{i_{{N_c} + 1}}, \cdots ,{i_N}} \!\right\}}^{N - {N_c}} \!{{{\left(\! { - 1} \!\right)}^{k'}}\!\sum\limits_{{j_1} = {j_0} + {N_c} + 1}^{N - k' + 1}\! { \cdots \!\sum\limits_{{j_{k'}} = {j_{k' - 1}} + 1}^N \!{\frac{1}{{\left(\! {\sum\limits_{l = 1}^{{N_c}} {\frac{1}{{{{\bar \gamma }_{{i_l}}}}}\! + \!\sum\limits_{m = 1}^{k'}\! {\frac{1}{{{{\bar \gamma }_{{i_{{j_m}}}}}}} - \frac{{{N_c}}}{{{{\bar \gamma }_{{i_q}}}}}} } } \!\right)}}\left\{\! {\exp \!\left( \!{ - \!\frac{z}{{{{\bar \gamma }_{{i_q}}}}}} \!\right) - \exp \!\left(\! { - \!\left( \!{\sum\limits_{l = 1}^{{N_c}} \!{\frac{1}{{{{\bar \gamma }_{{i_l}}}}} \!+ \!\sum\limits_{m = 1}^{k'}\! {\frac{1}{{{{\bar \gamma }_{{i_{{j_m}}}}}}}} } } \!\right)\frac{z}{{{N_c}}}} \!\right)} \!\right\}} } } } \!\right].
\end{aligned}
\end{equation}
Finally, by simply applying a basic exponential integration \cite{kn:abramowitz} over $z_2$ after substituting (\ref{eq:23}) in (\ref{eq:11}) and then replacing $z$ in (\ref{eq:23}) by $z_2$, we can obtain the target CDF in closed-form as shown in (\ref{eq:14}).

Note that, in the case of the closed-form expression of the CDF of the ${N_s}/{N}$-GSC output SNR given in (\ref{eq:19}), we can directly apply the same approach for (\ref{eq:13}) just by replacing $N_s$ with $N_c$.

\section{Joint PDF of   $Y$ and $W_1$} \label{appendix_3}
In this case, the target 2-dimensional joint PDF of $Y$ and $W_1$ can be obtained starting from the 4-dimensional joint PDF of  $Z_1  = \sum\limits_{i = 1}^{{N_c-N_s} - 1} {u_{i} }$, $Z_2  = u_{N_c-N_s}$, $Z_3 = \sum\limits_{i = {N_c-N_s} + 1}^{N_c  - 1} {u_{i }}$, and $Z_4  = u_{N_c}$  based on the proposed unified frame work in \cite{kn:unified_approach, kn:IND_MGF_sungsiknam_1}. As results, the target 2-dimensional joint PDF of interest, ${f_{Y,{W_1}}}\left( x , y \right)$, can be finally obtained from transformed higher dimensional joint PDFs as shown in (\ref{eq:12}).
Here, RVs, $Z_1$, $Z_2$, $Z_3$, and $Z_4$, have the following relationships
\begin{equation} \label{eq:relationship}
\overbrace {\underbrace {{u_{1}}, \cdots ,{u_{{N_c} - {N_s} - 1}}}_{{Z_1}},\underbrace {{u_{{N_c} - {N_s}}}}_{{Z_2}}}^Y,\overbrace {\underbrace {{u_{{N_c} - {N_s} + 1}}, \cdots ,{u_{{N_c} - 1}}}_{{Z_3}},\underbrace {{u_{{N_c}}}}_{{Z_4}}}^{{W_1}},{u_{{N_c} + 1}}, \cdots ,{u_{{N}}}.
\end{equation}
From (\ref{eq:relationship}), we can directly obtain the following valid conditions between these RVs  i) $Z_1 \ge \left(N_c - N_s -1\right) Z_2$, ii) $Z_3 \ge \left( N_s - 1 \right) Z_4$ and iii) $N_s \cdot Z_2 \ge Z_3 + Z_4$. For case i) and ii), by adding $Z_2$ to both sides of case i), we can obtain the following result as $Z_1+Z_2 \ge \left(N_c-N_s \right)Z_2$ while by adding $Z_4$ in case ii), we can obtain $Z_3+Z_4 \ge N_s \cdot Z_4$. Therefore, with the 4-dimensional joint PDF of $Z_1$, $Z_2$, $Z_3$, and $Z_4$, letting $X=Z_1+Z_2$ and $Y=Z_3+Z_4$, then we can obtain the target 2-dimensional joint PDF of $Z^{'}=[X,Y]$ by integrating over $z_2$ and $z_4$ yielding (\ref{eq:12}).

In (\ref{eq:12}), we now need to derive the 4-dimensional joint PDF of  $Z_1$, $Z_2$, $Z_3$, and $Z_4$. Fortunately, with the help of the derived result over i.n.d. exponential RVs in \cite[Eq. (54)]{kn:IND_MGF_sungsiknam_1}, we only need to evaluate the double integrations over $z_2$ and $z_4$. With \cite[Eq. (54)]{kn:IND_MGF_sungsiknam_1}, to evaluate the additional 2-fold integrations, the multiple product expression,  $\small\prod\limits_{j = {N_s} + 1}^N {\left( {1 - \exp \left( { - \frac{{{z_4}}}{{{{\bar \gamma }_{{i_j}}}}}} \right)} \right)}$, need to be converted to the summation expression. In this case, with the help of the property of exponential multiplication\footnote{The product of two exponential numbers of the same base can be simply represented as the sum of the exponents with the same base.}, this multiple product expression can be re-written as the following summation expression by adopting the derived result presented in (\ref{eq:AP_A_5})
\begin{equation} \small \label{eq:24}
\prod\limits_{j = {N_s} + 1}^N {\left( {1 - \exp \left( { - \frac{{{z_4}}}{{{{\bar \gamma }_{{i_j}}}}}} \right)} \right)}  = 1 + \sum\limits_{g = 1}^{N - {N_s}} {{{\left( { - 1} \right)}^g}\sum\limits_{{{j'}_1} = {{j'}_0} + {N_s} + 1}^{N - g + 1} { \cdots \sum\limits_{{{j'}_g} = {{j'}_{g - 1}} + 1}^N {\exp \left( { - \sum\limits_{m = 1}^g {\frac{{{z_4}}}{{{{\bar \gamma }_{{i_{{{j'}_m}}}}}}}} } \right)} } }.
\end{equation}
Then, inserting the re-written expression of \cite[Eq. (54)]{kn:IND_MGF_sungsiknam_1} as the summation expression into (\ref{eq:12}) and then after some manipulations, (\ref{eq:12}) can be re-written as

\begin{equation} \scriptsize \label{eq:arxiv_1}
\!\!\!\!\!\!\!\!\!\!\!\!\!\begin{aligned}
&{f_{Y,{W_1}}}\!\left(\! {x,y} \!\right)
\\
&= \!\sum\limits_{\scriptstyle {i_{{N_c}}}, \ldots ,{i_{N_1}} \atop
  \scriptstyle {i_{{N_c}}} \ne  \cdots  \ne {i_{N_1}}}^{1,2, \ldots ,{N_1}} \! {\frac{1}{{{{\bar \gamma }_{{i_{{N_c}}}}}}}\! \sum\limits_{\scriptstyle {i_{N_c - N_s}} = 1 \atop
  \scriptstyle {i_{N_c - N_s}} \ne {i_{{N_c}, \ldots ,}}{i_{N_1}}}^{N_1}\! {\frac{1}{{{{\bar \gamma }_{{i_{N_c - N_s}}}}}} \! \sum\limits_{\left\{\! {{i_{{N_c - N_s} + 1}}, \ldots ,{i_{{N_c} - 1}}} \!\right\} \in {{\mathop{\rm P}\nolimits} _{{ N_s} \!- \!1}}\left(\! {{I_{N_1}} \!-\! \left\{\! {{i_{N_c - N_s}}} \!\right\} \!- \!\left\{\! {{i_{{N_c}}}, \ldots ,{i_{N_1}}} \!\right\}} \!\right)} \! {\sum\limits_{k = {N_c - N_s} + 1}^{{N_c} - 1}\! {{C_{k,{N_c - N_s} + 1,{N_c} - 1}}} } } }
\\
&\left[ {\sum\limits_{\left\{\! {{i_1}, \ldots ,{i_{{N_c - N_s} - 1}}} \!\right\} \in {{\mathop{\rm P}\nolimits} _{{N_c - N_s} - 1}}\!\left(\! {{I_{N_1}} - \left\{\! {{i_{N_c - N_s}}} \!\right\} - \left\{\! {{i_{{N_c}}}, \ldots ,{i_{N_1}}} \!\right\} - \left\{\! {{i_{{N_c - N_s} + 1}}, \ldots ,{i_{{N_c} - 1}}} \!\right\}} \!\right)} \!{\sum\limits_{h = 1}^{{N_c - N_s} - 1} \!{{C_{h,1,{N_c - N_s} - 1}}} } \!\exp \!\left(\! { - \!\frac{x}{{{{\bar \gamma }_{{i_h}}}}}} \!\right)\!\exp \!\left(\! { - \!\frac{y}{{{{\bar \gamma }_{{i_k}}}}}} \!\right)} \right.
\\
&\quad\times \!\int_0^{\frac{y}{{{N_s}}}}\! {\int_{\frac{y}{{{ N_s}}}}^{\frac{x}{{N_c - N_s}}}\! {\exp \!\left( \!{ -\! \left(\! {\sum\limits_{l = 1}^{N_c - N_s}\! {\left(\! {\frac{1}{{{{\bar \gamma }_{{i_l}}}}}} \!\right) \!- \!\frac{N_c-N_s}{{{{\bar \gamma }_{{i_h}}}}}} } \!\right){z_2}} \!\right)\!\exp \!\left(\! { - \!\left( \!{\sum\limits_{l = {N_c - N_s} + 1}^{{N_c}}\! {\left( \!{\frac{1}{{{{\bar \gamma }_{{i_l}}}}}} \!\right)\! - \!\frac{{\left(\! {{ N_s}} \!\right)}}{{{{\bar \gamma }_{{i_k}}}}}} } \!\right){z_4}} \!\right)U\!\left( \!{x \!- \!\left({N_c \!- \!N_s}\right) \cdot {z_2}} \!\right)\!U\!\left(\! {y \!- \!\left(\! {{N_s}} \!\right) \cdot {z_4}} \!\right)d{z_2}d{z_4}} }
\\
& + \!\sum\limits_{l = 1}^{{ N_s} - 1} \!{{{{\left(\! { - \!1} \!\right)}^l}\!\sum\limits_{{j_1} = {j_0} + {N_c - N_s} + 1}^{{N_c} - l} \!{ \cdots \sum\limits_{{j_l} = {j_{l - 1}} + 1}^{{N_c} - 1}\! {} } }  \!\sum\limits_{\left\{ \!{{i_1}, \ldots ,{i_{{N_c - N_s} - 1}}} \!\right\} \in {{\mathop{\rm P}\nolimits} _{{N_c - N_s} - 1}}\!\left(\! {{I_{N_1}} \!- \!\left\{ \!{{i_{N_c - N_s}}} \!\right\} \!- \!\left\{\! {{i_{{N_c}}}, \ldots ,{i_{N_1}}} \!\right\} \!- \!\left\{\! {{i_{{N_c - N_s} + 1}}, \ldots ,{i_{{N_c} - 1}}} \!\right\}} \!\right)}   }
\\
&\quad\sum\limits_{h = 1}^{{N_c - N_s} - 1} {{C_{h,1,{N_c - N_s} - 1}}} \! \exp \!\left( \!{ -\! \frac{x}{{{{\bar \gamma }_{{i_h}}}}}} \!\right)\!\exp \!\left( \!{ -\! \frac{y}{{{{\bar \gamma }_{{i_k}}}}}} \!\right)
\\
&\quad\times \!\int_0^{\frac{y}{{{N_s}}}} \!{\int_{\frac{y}{{{N_s}}}}^{\frac{x}{{N_c - N_s}}}\! {\exp \!\left(\! { - \!\left( \!{\sum\limits_{l = 1}^{N_c - N_s} \!{\left(\! {\frac{1}{{{{\bar \gamma }_{{i_l}}}}}} \!\right) \!+ \!\sum\limits_{q = 1}^l\! {\frac{1}{{{{\bar \gamma }_{{i_{{j_q}}}}}}}}  \!-\! \frac{{N_c - N_s}}{{{{\bar \gamma }_{{i_h}}}}} \!- \!\frac{l}{{{{\bar \gamma }_{{i_k}}}}}} }\! \right)\!{z_2}} \!\right)\!\exp \!\left(\! { - \!\left(\! {\sum\limits_{l = {N_c - N_s} + 1}^{{N_c}} \!{\left( \!{\frac{1}{{{{\bar \gamma }_{{i_l}}}}}} \!\right) \!- \!\frac{{\left( \!{{ N_s} - l} \!\right)}}{{{{\bar \gamma }_{{i_k}}}}} \!- \!\sum\limits_{q = 1}^l \!{\frac{1}{{{{\bar \gamma }_{{i_{{j_q}}}}}}}} } } \!\right)\!{z_4}} \!\right)} }
\\
&\left. \sizecorr{\left[ {\sum\limits_{\left\{ {{i_1}, \ldots ,{i_{{N_c - N_s} - 1}}} \right\} \in {{\mathop{\rm P}\nolimits} _{{N_c - N_s} - 1}}\left( {{I_{N_1}} - \left\{ {{i_{N_c - N_s}}} \right\} - \left\{ {{i_{{N_c}}}, \ldots ,{i_{N_1}}} \right\} - \left\{ {{i_{{N_c - N_s} + 1}}, \ldots ,{i_{{N_c} - 1}}} \right\}} \right)} {\sum\limits_{h = 1}^{{N_c - N_s} - 1} {{C_{h,1,{N_c - N_s} - 1}}} } \exp \left( { - \frac{x}{{{{\bar \gamma }_{{i_h}}}}}} \right)\exp \left( { - \frac{y}{{{{\bar \gamma }_{{i_k}}}}}} \right)} \right.}
\quad\quad U\left( {x - \left({N_c \!- \!N_s}\right) \cdot {z_2}} \right)U\left( {y - \left( {l \cdot {z_2} + \left( {{N_s} - l} \right) \cdot {z_4}} \right)} \right)d{z_2}d{z_4} \right]
\\
&{ + \sum\limits_{\scriptstyle {i_{{N_c}}}, \ldots ,{i_{N_1}} \atop
  \scriptstyle {i_{{N_c}}} \ne  \cdots  \ne {i_{N_1}}}^{1,2, \ldots ,{N_1}} {\frac{1}{{{{\bar \gamma }_{{i_{{N_c}}}}}}}\sum\limits_{g = 1}^{{N_1} - {N_c}} {{{\left( { - 1} \right)}^g}\sum\limits_{{{j'}_1} = {{j'}_0} + {N_c} + 1}^{{N_1} - g + 1} { \cdots \sum\limits_{{{j'}_g} = {{j'}_{g - 1}} + 1}^{N_1} {\sum\limits_{\scriptstyle {i_{N_c - N_s}} = 1 \atop
  \scriptstyle {i_{N_c - N_s}} \ne {i_{{N_c}, \ldots ,}}{i_{N_1}}}^{N_1} {\frac{1}{{{{\bar \gamma }_{{i_{N_c - N_s}}}}}} } } } } } }
\\
&\sum\limits_{\left\{ {{i_{{N_c - N_s} + 1}}, \ldots ,{i_{{N_c} - 1}}} \right\} \in {{\mathop{\rm P}\nolimits} _{{ N_s} - 1}}\left( {{I_{N_1}} - \left\{ {{i_{N_c - N_s}}} \right\} - \left\{ {{i_{{N_c}}}, \ldots ,{i_{N_1}}} \right\}} \right)} {\sum\limits_{k = {N_c - N_s} + 1}^{{N_c} - 1} {{C_{k,{N_c - N_s} + 1,{N_c} - 1}}} }
\\
&\left[ {\sum\limits_{\left\{\! {{i_1}, \ldots ,{i_{{N_c - N_s} - 1}}}\! \right\} \in {{\mathop{\rm P}\nolimits} _{{N_c - N_s} - 1}}\!\left(\! {{I_{N_1}} - \left\{\! {{i_{N_c - N_s}}} \!\right\} - \left\{\! {{i_{{N_c}}}, \ldots ,{i_{N_1}}}\! \right\} - \left\{\! {{i_{{N_c - N_s} + 1}}, \ldots ,{i_{{N_c} - 1}}}\! \right\}} \!\right)} \!{\sum\limits_{h = 1}^{{N_c - N_s} - 1}\! {{C_{h,1,{N_c - N_s} - 1}}} } \!\exp \!\left(\! { -\! \frac{x}{{{{\bar \gamma }_{{i_h}}}}}} \!\right)\!\exp \!\left(\! { - \!\frac{y}{{{{\bar \gamma }_{{i_k}}}}}} \!\right)}\right.
\\
&\quad\times \int_0^{\frac{y}{{{N_s}}}} {\int_{\frac{y}{{{N_s}}}}^{\frac{x}{{N_c - N_s}}} {\exp \left( { - \left( {\sum\limits_{l = 1}^{N_c - N_s} {\left( {\frac{1}{{{{\bar \gamma }_{{i_l}}}}}} \right) - \frac{{N_c - N_s}}{{{{\bar \gamma }_{{i_h}}}}}} } \right){z_2}} \right)\exp \left( { - \left( {\sum\limits_{l = {N_c - N_s} + 1}^{{N_c}} {\left( {\frac{1}{{{{\bar \gamma }_{{i_l}}}}}} \right) + \sum\limits_{m = 1}^g {\frac{1}{{{{\bar \gamma }_{{i_{{{j'}_m}}}}}}}}  - \frac{{\left( {{N_s}} \right)}}{{{{\bar \gamma }_{{i_k}}}}}} } \right){z_4}} \right)} }
\\
&\quad\quad U\left( {x - \left({N_c \!- \!N_s}\right) \cdot {z_2}} \right)U\left( {y - \left( {{N_s}} \right) \cdot {z_4}} \right)d{z_2}d{z_4}
\\
&+ \!\sum\limits_{l = 1}^{ {N_s} - 1} \!{ {{{\left(\! { - \!1} \!\right)}^l}\!\sum\limits_{{j_1} = {j_0} + {N_c - N_s} + 1}^{{N_c} - l}\! { \cdots \!\sum\limits_{{j_l} = {j_{l - 1}} + 1}^{{N_c} - 1}\! {} } } \!\sum\limits_{\left\{\! {{i_1}, \ldots ,{i_{{N_c - N_s} - 1}}} \!\right\} \in {{\mathop{\rm P}\nolimits} _{{N_c - N_s} - 1}}\!\left(\! {{I_{N_1}} \!- \!\left\{\! {{i_{N_c - N_s}}} \!\right\} \!- \!\left\{\! {{i_{{N_c}}}, \ldots ,{i_{N_1}}} \!\right\} \!-\! \left\{\! {{i_{{N_c - N_s} + 1}}, \ldots ,{i_{{N_c} - 1}}} \!\right\}} \!\right)}  }
\\
&\quad {\sum\limits_{h = 1}^{{N_c - N_s} - 1} {{C_{h,1,{N_c - N_s} - 1}}}} \exp \left( { - \frac{x}{{{{\bar \gamma }_{{i_h}}}}}} \right)\exp \left( { - \frac{y}{{{{\bar \gamma }_{{i_k}}}}}} \right)
\\
& \quad { \times \!\int_0^{\frac{y}{{{N_s}}}} \!{\int_{\frac{y}{{{N_s}}}}^{\frac{x}{{N_c - N_s}}} \! {\exp \!\left( \!{ - \!\left(\! {\sum\limits_{l = 1}^{N_c - N_s}\! {\left(\! {\frac{1}{{{{\bar \gamma }_{{i_l}}}}}}\! \right)\! + \!\sum\limits_{q = 1}^l \!{\frac{1}{{{{\bar \gamma }_{{i_{{j_q}}}}}}}} \! - \! \frac{{N_c - N_s}}{{{{\bar \gamma }_{{i_h}}}}} \!- \!\frac{l}{{{{\bar \gamma }_{{i_k}}}}}} } \!\right)\!{z_2}} \!\right)\!\exp\! \left(\! { - \!\left(\! {\sum\limits_{l = {N_c - N_s} + 1}^{{N_c}} \!{\left( \!{\frac{1}{{{{\bar \gamma }_{{i_l}}}}}} \!\right) \!+ \!\sum\limits_{m = 1}^g \!{\frac{1}{{{{\bar \gamma }_{{i_{{{j'}_m}}}}}}}} \! -\! \frac{{\left(\! {{ N_s} - l} \!\right)}}{{{{\bar \gamma }_{{i_k}}}}} \!- \!\sum\limits_{q = 1}^l \!{\frac{1}{{{{\bar \gamma }_{{i_{{j_q}}}}}}}} } } \!\right)\!{z_4}}\! \right)} } }
\\
&\left. \sizecorr{\left[ {\sum\limits_{\left\{ {{i_1}, \ldots ,{i_{{N_c - N_s} - 1}}} \right\} \in {{\mathop{\rm P}\nolimits} _{{N_c - N_s} - 1}}\left( {{I_{N_1}} - \left\{ {{i_{N_c - N_s}}} \right\} - \left\{ {{i_{{N_c}}}, \ldots ,{i_{N_1}}} \right\} - \left\{ {{i_{{N_c - N_s} + 1}}, \ldots ,{i_{{N_c} - 1}}} \right\}} \right)} {\sum\limits_{h = 1}^{{N_c - N_s} - 1} {{C_{h,1,{N_c - N_s} - 1}}} } \exp \left( { - \frac{x}{{{{\bar \gamma }_{{i_h}}}}}} \right)\exp \left( { - \frac{y}{{{{\bar \gamma }_{{i_k}}}}}} \right)}\right.}
\quad\quad U\left( {x - \left({N_c \!- \!N_s}\right) \cdot {z_2}} \right)U\left( {y - \left( {l \cdot {z_2} + \left( {{N_s} - l} \right) \cdot {z_4}} \right)} \right)d{z_2}d{z_4} \right].
\end{aligned}
\end{equation}

With (\ref{eq:arxiv_1}), we now need to evaluate the following four integral terms
\\
\noindent A) For the first integral term:
\begin{equation}  \small  \label{eq:25}
\begin{aligned}
&\int_0^{\frac{y}{{N_s}}}\! {\int_{\frac{y}{{ {N_s}}}}^{\frac{x}{{N_c-N_s}}}\! {\exp \!\left( \!{ - \!\left( \!{\sum\limits_{l = 1}^{N_c-N_s}\! {\left(\! {\frac{1}{{{{\bar \gamma }_{{i_l}}}}}}\! \right)\! - \!\frac{{N_c\!-\!N_s}}{{{{\bar \gamma }_{{i_h}}}}}} }\! \right)\!{z_2}}\! \right)} }
\\
&\quad \exp \!\left(\! { - \!\left( \!{\sum\limits_{l = {N_c-N_s} + 1}^{{N_c}}\! {\left(\! {\frac{1}{{{{\bar \gamma }_{{i_l}}}}}} \!\right)\! -\! \frac{{\left(\! { {N_s}} \!\right)}}{{{{\bar \gamma }_{{i_k}}}}}} } \!\right)\!{z_4}} \!\right)\!U\!\left( \!{x \!- \!\left(\!{N_c\!-\!N_s}\!\right) \cdot {z_2}} \!\right)\!U\!\left( \!{y \!- \!\left(\! {{N_s}} \!\right) \cdot {z_4}} \!\right)\!d{z_2}d{z_4},
\end{aligned}
\end{equation}
\noindent B) For the second integral term:
\begin{equation}  \small  \label{eq:26}
\begin{aligned}
&\int_0^{\frac{y}{{{N_s}}}} \!{\int_{\frac{y}{{ {N_s}}}}^{\frac{x}{{N_c-N_s}}} \!{\exp\! \left(\! { -\! \left(\! {\sum\limits_{l = 1}^{N_c-N_s}\! {\left(\! {\frac{1}{{{{\bar \gamma }_{{i_l}}}}}}\! \right)\! + \!\sum\limits_{q = 1}^l \! {\frac{1}{{{{\bar \gamma }_{{i_{{j_q}}}}}}}} \! - \!\frac{{N_c\!-\!N_s}}{{{{\bar \gamma }_{{i_h}}}}} \!- \!\frac{l}{{{{\bar \gamma }_{{i_k}}}}}} } \!\right)\!{z_2}} \!\right)} }
\\
&\quad\exp \!\left(\! { - \!\left( \!{\sum\limits_{l = {N_c-N_s} + 1}^{{N_c}}\! {\left(\! {\frac{1}{{{{\bar \gamma }_{{i_l}}}}}} \!\right) \!- \!\frac{{\left( \!{{N_s} \!- \!l} \!\right)}}{{{{\bar \gamma }_{{i_k}}}}} \!- \!\sum\limits_{q = 1}^l \!{\frac{1}{{{{\bar \gamma }_{{i_{{j_q}}}}}}}} } } \!\right)\!{z_4}} \!\right) U\!\left( \!{x \!- \!\left(\!{N_c\!-\!N_s}\!\right) \cdot {z_2}}\! \right)\!U\!\left( \!{y \!- \!\left( \!{l \cdot {z_2} \!+ \!\left( \!{{N_s} \!- \!l} \!\right) \cdot {z_4}} \!\right)} \!\right)d{z_2}d{z_4},
\end{aligned}
\end{equation}
\noindent C) For the third integral term:
\begin{equation} \small \label{eq:27}
\begin{aligned}
&\int_0^{\frac{y}{{{N_s}}}}\! {\int_{\frac{y}{{{N_s}}}}^{\frac{x}{{N_c-N_s}}} \!{\exp\! \left(\! { - \!\left(\! {\sum\limits_{l = 1}^{N_c-N_s} \!{\left( \!{\frac{1}{{{{\bar \gamma }_{{i_l}}}}}} \!\right) \!- \!\frac{{N_c\!-\!N_s}}{{{{\bar \gamma }_{{i_h}}}}}} } \!\right){z_2}} \!\right)\!\exp\! \left(\! { - \!\left(\! {\sum\limits_{l = {N_c-N_s} + 1}^{{N_c}} \!{\left( \!{\frac{1}{{{{\bar \gamma }_{{i_l}}}}}} \!\right) \!+\! \sum\limits_{m = 1}^g \!{\frac{1}{{{{\bar \gamma }_{{i_{{{j'}_m}}}}}}}} \! - \!\frac{{\left(\! {{N_s}}\! \right)}}{{{{\bar \gamma }_{{i_k}}}}}} } \!\right){z_4}} \!\right)} }
\\
&\quad U\left( {x - \left({N_c-N_s}\right)\cdot {z_2}} \right)U\left( {y - \left( {{N_s}} \right) \cdot {z_4}} \right)d{z_2}d{z_4},
\end{aligned}
\end{equation}
\noindent D) For the fourth integral term:
\begin{equation} \small  \label{eq:28}
\!\!\!\begin{aligned}
&\int_0^{\frac{y}{{{N_s}}}} \!{\int_{\frac{y}{{{N_s}}}}^{\frac{x}{{N_c-N_s}}} \!{\exp \!\left(\! { - \!\left(\! {\sum\limits_{l = 1}^{N_c-N_s} \!{\left(\! {\frac{1}{{{{\bar \gamma }_{{i_l}}}}}} \!\right) \!+ \!\sum\limits_{q = 1}^l \!{\frac{1}{{{{\bar \gamma }_{{i_{{j_q}}}}}}}} \! - \! \frac{{N_c\!-\!N_s}}{{{{\bar \gamma }_{{i_h}}}}} \!-\! \frac{l}{{{{\bar \gamma }_{{i_k}}}}}} } \!\right)\!{z_2}} \!\right)} }
\\
&\quad\exp \!\left( \!{ -\! \left( \!{\sum\limits_{l = {N_c-N_s} + 1}^{{N_c}} \!{\left(\! {\frac{1}{{{{\bar \gamma }_{{i_l}}}}}} \!\right) \!+ \!\sum\limits_{m = 1}^g \!{\frac{1}{{{{\bar \gamma }_{{i_{{{j'}_m}}}}}}}} \! - \!\frac{{\left( \!{ {N_s} \!-\! l} \!\right)}}{{{{\bar \gamma }_{{i_k}}}}} \!- \!\sum\limits_{q = 1}^l \!{\frac{1}{{{{\bar \gamma }_{{i_{{j_q}}}}}}}} } } \!\right)\!{z_4}} \!\right) U\left( {x - \left({N_c-N_s}\right) \cdot {z_2}} \right)
\\
&\quad U\left( {y - \left( {l \cdot {z_2} + \left( {{N_s} - l} \right) \cdot {z_4}} \right)} \right)d{z_2}d{z_4}.
\end{aligned}
\end{equation}
\normalsize

For the first and the third integral terms, closed-form expression can be obtained by simply applying basic exponential integration \cite{kn:abramowitz}, using the following useful common function
\begin{equation} \label{eq:common_1}
\small
\begin{array}{l}
{\rm I}\left( {x,e,a,b;y,f,c,d} \right)
 = \int_c^d {\int_a^b {\exp \left( {e \cdot x} \right)\exp \left( {f \cdot y} \right)dxdy} } \\
 = \frac{1}{{e \cdot f}}\left\{ {\exp \left( {e \cdot b} \right) - \exp \left( {e \cdot a} \right)} \right\}\left\{ {\exp \left( {f \cdot d} \right) - \exp \left( {f \cdot c} \right)} \right\}.
\end{array}
\end{equation}
With (\ref{eq:common_1}), by letting \small$\alpha  =  - \left( {\sum\limits_{l = {N_c-N_s} + 1}^{{N_c}} {\left( {\frac{1}{{{{\bar \gamma }_{{i_l}}}}}} \right) - \frac{{\left( {{N_s}} \right)}}{{{{\bar \gamma }_{{i_k}}}}}} } \right)$\normalsize, \small$\alpha ' =  - \left( {\sum\limits_{l = {N_c-N_s} + 1}^{{N_c}} {\left( {\frac{1}{{{{\bar \gamma }_{{i_l}}}}}} \right) + \sum\limits_{m = 1}^g {\frac{1}{{{{\bar \gamma }_{{i_{{{j'}_m}}}}}}}}  - \frac{{\left( {{N_s}} \right)}}{{{{\bar \gamma }_{{i_k}}}}}} } \right)$\normalsize, and \small$\beta  =  - \left( {\sum\limits_{l = 1}^{{N_c-N_s}} {\left( {\frac{1}{{{{\bar \gamma }_{{i_l}}}}}} \right) - \frac{{N_c-N_s}}{{{{\bar \gamma }_{{i_h}}}}}} } \right)$\normalsize, the closed-form expression of the first integral term can be obtained by simply applying the basic exponential integration \cite{kn:abramowitz} as
\begin{equation} \label{eq:29}
\small
\begin{array}{l}
\int_0^{\frac{y}{{{N_s}}}} {\int_{\frac{y}{{{N_s}}}}^{\frac{x}{{{N_c} - {N_s}}}} {\exp \left( {\beta {z_2}} \right)\exp \left( {\alpha {z_4}} \right)U\left( {x - \left( {{N_c} - {N_s}} \right) \cdot {z_2}} \right)U\left( {y - \left( {{N_s}} \right) \cdot {z_4}} \right)d{z_2}d{z_4}} } \\
 = \int_0^{\frac{y}{{{N_s}}}} {\int_{\frac{y}{{{N_s}}}}^{\frac{x}{{{N_c} - {N_s}}}} {\exp \left( {\beta {z_2}} \right)\exp \left( {\alpha {z_4}} \right)d{z_2}d{z_4}} } \\
 = {\rm I}\left( {{z_2},\beta ,\frac{y}{{{N_s}}},\frac{x}{{{N_c} - {N_s}}};{z_4},\alpha ,0,\frac{y}{{{N_s}}}} \right).
\end{array}
\end{equation}
Similarly, for the third integral term, we can also obtain closed-form expressions simply by replacing $\alpha$ with $\alpha '$ on (\ref{eq:29}) as
\begin{equation} \label{eq:30}
\small
\begin{array}{l}
\int_0^{\frac{y}{{{N_s}}}} {\int_{\frac{y}{{{N_s}}}}^{\frac{x}{{{N_c} - {N_s}}}} {\exp \left( {\beta {z_2}} \right)\exp \left( {\alpha '{z_4}} \right)U\left( {x - \left( {{N_c} - {N_s}} \right) \cdot {z_2}} \right)U\left( {y - \left( {{N_s}} \right) \cdot {z_4}} \right)d{z_2}d{z_4}} } \\
 = \int_0^{\frac{y}{{{N_s}}}} {\int_{\frac{y}{{{N_s}}}}^{\frac{x}{{{N_c} - {N_s}}}} {\exp \left( {\beta {z_2}} \right)\exp \left( {\alpha '{z_4}} \right)d{z_2}d{z_4}} } \\
 = {\rm I}\left( {{z_2},\beta ,\frac{y}{{{N_s}}},\frac{x}{{{N_c} - {N_s}}};{z_4},\alpha ',0,\frac{y}{{{N_s}}}} \right).
\end{array}
\end{equation}

However, for the second and the fourth integral terms, we need to consider two cases separately based on the valid integration region of $z_2$. More specifically, $z_2$ should satisfy two following conditions as i) $z_2 \le \frac{x}{{N_c-N_s}}$ and ii) $z_2 \le \frac{y-\left( N_s-l \right)z_4}{l}$ and it leads $z_2 \le {\rm min}\left[ \frac{x}{{N_c-N_s}}, \frac{y-\left( {N_s}-l \right)z_4}{l}\right]$. Based on it, if $\frac{x}{{N_c-N_s}} \le \frac{y-\left( {N_s}-l \right)z_4}{l}$, then the valid integral regions for $z_2$ is unchanged, $\frac{y}{{N_s}} <z_2< \frac{x}{{N_c-N_s}}$, but the valid integral regions for $z_4$ becomes changed to $0 < z_4 < {\rm min}\left[\frac{y}{{N_s}}, \frac{y-\frac{l}{{N_c-N_s}}x}{\left({N_s}-l\right)}\right]$ from $0 < z_4 < \frac{y}{{N_s}}$. Otherwise, the valid integral regions for $z_2$ is changed as $\frac{y}{{N_s}} <z_2< \frac{y-\left( {N_s}-l \right)z_4}{l}$ and for $z_4$, we need to consider both cases of $0 < z_4 < {\rm min}\left[\frac{y}{{N_s}}, \frac{y-\frac{l}{{N_c-N_s}}x}{\left({N_s}-l\right)}\right]$ and $0 < z_4 < \frac{y}{{N_s}}$ by considering the unit step function, ${\left\{ {1 - U\left( {\frac{x}{{N_c-N_s}} - \frac{{y - \left( {{N_s} - l} \right) \cdot {z_4}}}{l}} \right)} \right\}}$. As results, the second and the fourth integral terms can be re-written, respectively, as

\noindent For the second integral term:
\begin{equation} \small \label{eq:arxiv_2}
\begin{aligned}
&\int_0^{\min \left[ {\frac{y}{{{N_s}}},\frac{{y - \frac{l}{{N_c-N_s}} \cdot x}}{{\left( {{N_s} - l} \right)}}} \right]} {\exp \left( { - \left( {\sum\limits_{l = {N_c-N_s} + 1}^{{N_c}} {\left( {\frac{1}{{{{\bar \gamma }_{{i_l}}}}}} \right) - \frac{{\left( {{N_s} - l} \right)}}{{{{\bar \gamma }_{{i_k}}}}} - \sum\limits_{q = 1}^l {\frac{1}{{{{\bar \gamma }_{{i_{{j_q}}}}}}}} } } \right){z_4}} \right) }
\\
&\int_{\frac{y}{{{N_s}}}}^{\frac{x}{{N_c-N_s}}} {\exp \left( { - \left( {\sum\limits_{l = 1}^{N_c-N_s} {\left( {\frac{1}{{{{\bar \gamma }_{{i_l}}}}}} \right) + \sum\limits_{q = 1}^l {\frac{1}{{{{\bar \gamma }_{{i_{{j_q}}}}}}}}  - \frac{{N_c-N_s}}{{{{\bar \gamma }_{{i_h}}}}} - \frac{l}{{{{\bar \gamma }_{{i_k}}}}}} } \right){z_2}} \right)d{z_2}d{z_4}}
\\
&+ \left\{ {\int_0^{\frac{y}{{{N_s}}}} {\exp \left( { - \left( {\sum\limits_{l = {N_c-N_s} + 1}^{{N_c}} {\left( {\frac{1}{{{{\bar \gamma }_{{i_l}}}}}} \right) - \frac{{\left( {{N_s} - l} \right)}}{{{{\bar \gamma }_{{i_k}}}}} - \sum\limits_{q = 1}^l {\frac{1}{{{{\bar \gamma }_{{i_{{j_q}}}}}}}} } } \right){z_4}} \right) } } \right.
\\
&\quad\quad\int_{\frac{y}{{{N_s}}}}^{\frac{{y - \left( {{N_s} - l} \right) \cdot {z_4}}}{l}} {\exp \left( { - \left( {\sum\limits_{l = 1}^{N_c-N_s} {\left( {\frac{1}{{{{\bar \gamma }_{{i_l}}}}}} \right) + \sum\limits_{q = 1}^l {\frac{1}{{{{\bar \gamma }_{{i_{{j_q}}}}}}}}  - \frac{N_c-N_s}{{{{\bar \gamma }_{{i_h}}}}} - \frac{l}{{{{\bar \gamma }_{{i_k}}}}}} } \right){z_2}} \right)d{z_2}d{z_4}}
\\
& \quad\quad{ - \int_0^{\min \left[ {\frac{y}{{{N_s}}},\frac{{y - \frac{l}{{N_c-N_s}} \cdot x}}{{\left( {{N_s} - l} \right)}}} \right]} {\exp \left( { - \left( {\sum\limits_{l = {N_c-N_s} + 1}^{{N_c}} {\left( {\frac{1}{{{{\bar \gamma }_{{i_l}}}}}} \right) - \frac{{\left( {{N_s} - l} \right)}}{{{{\bar \gamma }_{{i_k}}}}} - \sum\limits_{q = 1}^l {\frac{1}{{{{\bar \gamma }_{{i_{{j_q}}}}}}}} } } \right){z_4}} \right) } }
\\
&\quad\quad\quad\left. \sizecorr{+ \left\{ {\int_0^{\frac{y}{{{N_s}}}} {\exp \left( { - \left( {\sum\limits_{l = {N_c-N_s} + 1}^{{N_c}} {\left( {\frac{1}{{{{\bar \gamma }_{{i_l}}}}}} \right) - \frac{{\left( {{N_s} - l} \right)}}{{{{\bar \gamma }_{{i_k}}}}} - \sum\limits_{q = 1}^l {\frac{1}{{{{\bar \gamma }_{{i_{{j_q}}}}}}}} } } \right){z_4}} \right) } } \right.}
 \int_{\frac{y}{{{N_s}}}}^{\frac{{y - \left( {{N_s} - l} \right) \cdot {z_4}}}{l}} {\exp \left( { - \left( {\sum\limits_{l = 1}^{N_c-N_s} {\left( {\frac{1}{{{{\bar \gamma }_{{i_l}}}}}} \right) + \sum\limits_{q = 1}^l {\frac{1}{{{{\bar \gamma }_{{i_{{j_q}}}}}}}}  - \frac{N_c-N_s}{{{{\bar \gamma }_{{i_h}}}}} - \frac{l}{{{{\bar \gamma }_{{i_k}}}}}} } \right){z_2}} \right)d{z_2}d{z_4}} \right\},
\end{aligned}
\end{equation}
\noindent For the fourth integral term:
\begin{equation} \small \label{eq:arxiv_3}
\begin{aligned}
&\int_0^{\min \left[ {\frac{y}{{{N_s}}},\frac{{y - \frac{l}{{N_c-N_s}} \cdot x}}{{\left( {{N_s} - l} \right)}}} \right]} {\exp \left( { - \left( {\sum\limits_{l = {N_c-N_s} + 1}^{{N_c}} {\left( {\frac{1}{{{{\bar \gamma }_{{i_l}}}}}} \right) + \sum\limits_{m = 1}^g {\frac{1}{{{{\bar \gamma }_{{i_{{{j'}_m}}}}}}}}  - \frac{{\left( {{N_s} - l} \right)}}{{{{\bar \gamma }_{{i_k}}}}} - \sum\limits_{q = 1}^l {\frac{1}{{{{\bar \gamma }_{{i_{{j_q}}}}}}}} } } \right){z_4}} \right) }
\\
&\int_{\frac{y}{{{N_s}}}}^{\frac{x}{{N_c-N_s}}} {\exp \left( { - \left( {\sum\limits_{l = 1}^{N_c-N_s} {\left( {\frac{1}{{{{\bar \gamma }_{{i_l}}}}}} \right) + \sum\limits_{q = 1}^l {\frac{1}{{{{\bar \gamma }_{{i_{{j_q}}}}}}}}  - \frac{{N_c-N_s}}{{{{\bar \gamma }_{{i_h}}}}} - \frac{l}{{{{\bar \gamma }_{{i_k}}}}}} } \right){z_2}} \right)d{z_2}d{z_4}}
\\
&+ \left\{ {\int_0^{\frac{y}{{{N_s}}}} {\exp \left( { - \left( {\sum\limits_{l = {N_c-N_s} + 1}^{{N_c}} {\left( {\frac{1}{{{{\bar \gamma }_{{i_l}}}}}} \right) + \sum\limits_{m = 1}^g {\frac{1}{{{{\bar \gamma }_{{i_{{{j'}_m}}}}}}}}  - \frac{{\left( {{N_s} - l} \right)}}{{{{\bar \gamma }_{{i_k}}}}} - \sum\limits_{q = 1}^l {\frac{1}{{{{\bar \gamma }_{{i_{{j_q}}}}}}}} } } \right){z_4}} \right) } } \right.
\\
&\quad\quad\int_{\frac{y}{{{N_s}}}}^{\frac{{y - \left( {{N_s} - l} \right) \cdot {z_4}}}{l}} {\exp \left( { - \left( {\sum\limits_{l = 1}^{N_c-N_s} {\left( {\frac{1}{{{{\bar \gamma }_{{i_l}}}}}} \right) + \sum\limits_{q = 1}^l {\frac{1}{{{{\bar \gamma }_{{i_{{j_q}}}}}}}}  - \frac{{N_c-N_s}}{{{{\bar \gamma }_{{i_h}}}}} - \frac{l}{{{{\bar \gamma }_{{i_k}}}}}} } \right){z_2}} \right)d{z_2}d{z_4}}
\\
&\quad\quad{ - \int_0^{\min \left[ {\frac{y}{{{N_s}}},\frac{{y - \frac{l}{{N_c-N_s}} \cdot x}}{{\left( {{N_s} - l} \right)}}} \right]} {\exp \left( { - \left( {\sum\limits_{l = {N_c-N_s} + 1}^{{N_c}} {\left( {\frac{1}{{{{\bar \gamma }_{{i_l}}}}}} \right) + \sum\limits_{m = 1}^g {\frac{1}{{{{\bar \gamma }_{{i_{{{j'}_m}}}}}}}}  - \frac{{\left( {{N_s} - l} \right)}}{{{{\bar \gamma }_{{i_k}}}}} - \sum\limits_{q = 1}^l {\frac{1}{{{{\bar \gamma }_{{i_{{j_q}}}}}}}} } } \right){z_4}} \right) } }.
\\
&\quad\quad\quad\left. \int_{\frac{y}{{{N_s}}}}^{\frac{{y - \left( {{N_s} - l} \right) \cdot {z_4}}}{l}} {\exp \left( { - \left( {\sum\limits_{l = 1}^{N_c-N_s} {\left( {\frac{1}{{{{\bar \gamma }_{{i_l}}}}}} \right) + \sum\limits_{q = 1}^l {\frac{1}{{{{\bar \gamma }_{{i_{{j_q}}}}}}}}  - \frac{N_c-N_s}{{{{\bar \gamma }_{{i_h}}}}} - \frac{l}{{{{\bar \gamma }_{{i_k}}}}}} } \right){z_2}} \right)d{z_2}d{z_4}} \right\}.
\end{aligned}
\end{equation}

In (\ref{eq:arxiv_2}), the closed-form expression of the first double integral term can be obtained with the help of (\ref{eq:common_1}). However, for the second and the third double integral terms, the inner integral limit depends on the outer variable. Therefore, for these cases, we can obtain the closed-form expression with the help of another useful common function as
\begin{equation} \label{eq:common_2}
\small\begin{array}{l}
{\rm I}'\left( {x,e,a,b - b'y;y,f,c,d} \right)
 = \int_c^d {\exp \left( {f \cdot y} \right)\int_a^{b - b'y} {\exp \left( {e \cdot x} \right)dxdy} } \\
 = \frac{1}{e}\left[ {\exp \left( {e \cdot a} \right)\frac{1}{{\left( {f - e \cdot b'} \right)}}\left\{ {\exp \left( {\left( {f - e \cdot b'} \right) \cdot d} \right) - \exp \left( {\left( {f - e \cdot b'} \right) \cdot c} \right)} \right\}} \right.\\
\left. { - \exp \left( {e \cdot a} \right)\frac{1}{f}\left\{ {\exp \left( {f \cdot d} \right) - \exp \left( {f \cdot c} \right)} \right\}} \right].
\end{array}
\end{equation}
Then, by letting \small$\alpha '' =  - \left( {\sum\limits_{l = {N_c-N_s} + 1}^{{N_c}} {\left( {\frac{1}{{{{\bar \gamma }_{{i_{l}}}}}}} \right) - \frac{{\left( {{N_s} - l} \right)}}{{{{\bar \gamma }_{{i_{k}}}}}} - \sum\limits_{q = 1}^l {\frac{1}{{{{\bar \gamma }_{{i_{{j_q}}}}}}}} } } \right)$\normalsize and \small$\beta'  =  - \left( {\sum\limits_{l = 1}^{N_c-N_s} {\left( {\frac{1}{{{{\bar \gamma }_{{i_{l}}}}}}} \right) + \sum\limits_{q = 1}^l {\frac{1}{{{{\bar \gamma }_{{i_{{j_q}}}}}}}}  - \frac{{N_c-N_s}}{{{{\bar \gamma }_{{i_{h}}}}}} - \frac{l}{{{{\bar \gamma }_{{i_{k}}}}}}} } \right)$\normalsize, we can finally obtain the closed-form expression of (\ref{eq:arxiv_2}) as
\begin{equation} \label{eq:arxiv_4}
\small\begin{array}{l}
\left[ {{\rm I}\left( {{z_2},\beta ',\frac{y}{{{N_s}}},\frac{x}{{{N_c} - {N_s}}};{z_4},\alpha '',0,\min \left[ {\frac{y}{{{N_s}}},\frac{{y - \frac{l}{{{N_c} - {N_s}}} \cdot x}}{{\left( {{N_s} - l} \right)}}} \right]} \right)} \right.\\
\left. + {\rm I}'\left( {{z_2},\beta ',\frac{y}{{{N_s}}},\frac{{y - \left( {{N_s} - l} \right) \cdot {z_4}}}{l};{z_4},\alpha '',0,\frac{y}{{{N_s}}}} \right)
 { - {\rm I}'\left( {{z_2},\beta ',\frac{y}{{{N_s}}},\frac{{y - \left( {{N_s} - l} \right) \cdot {z_4}}}{l};{z_4},\alpha '',0,\min \left[ {\frac{y}{{{N_s}}},\frac{{y - \frac{l}{{{N_c} - {N_s}}} \cdot x}}{{\left( {{N_s} - l} \right)}}} \right]} \right)} \right].
\end{array}
\end{equation}

For (\ref{eq:arxiv_3}), by applying a similar approach and with the help of (\ref{eq:common_2}), we can also obtain the closed-form expression given in (\ref{eq:arxiv_3}). More specifically, we can also obtain the closed-form expressions simply by replacing $\alpha''$ with $\alpha '''$ on (\ref{eq:arxiv_4}) as
\begin{equation} \label{eq:arxiv_5}
\small\begin{array}{l}
\left[ {{\rm I}\left( {{z_2},\beta ',\frac{y}{{{N_s}}},\frac{x}{{{N_c} - {N_s}}};{z_4},\alpha ''',0,\min \left[ {\frac{y}{{{N_s}}},\frac{{y - \frac{l}{{{N_c} - {N_s}}} \cdot x}}{{\left( {{N_s} - l} \right)}}} \right]} \right)} \right.\\
\left. + {\rm I}'\left( {{z_2},\beta ',\frac{y}{{{N_s}}},\frac{{y - \left( {{N_s} - l} \right) \cdot {z_4}}}{l};{z_4},\alpha ''',0,\frac{y}{{{N_s}}}} \right)
 { - {\rm I}'\left( {{z_2},\beta ',\frac{y}{{{N_s}}},\frac{{y - \left( {{N_s} - l} \right) \cdot {z_4}}}{l};{z_4},\alpha ''',0,\min \left[ {\frac{y}{{{N_s}}},\frac{{y - \frac{l}{{{N_c} - {N_s}}} \cdot x}}{{\left( {{N_s} - l} \right)}}} \right]} \right)} \right],
\end{array}
\end{equation}
where \small$\alpha ''' =  - \Bigg( \sum\limits_{l = {N_c-N_s} + 1}^{N_c} \left( \frac{1}{{\bar \gamma }_{i_{l}}} \right) + \sum\limits_{m = 1}^g {\frac{1}{{{{\bar \gamma }_{{i_{{{j'}_m}}}}}}}}  - \frac{{\left( {{N_s} - l} \right)}}{{{{\bar \gamma }_{{i_{k}}}}}} - \sum\limits_{q = 1}^l {\frac{1}{{{{\bar \gamma }_{{i_{{j_q}}}}}}}}   \Bigg)$\normalsize.

Then, by substituting all these closed-form results and after a few manipulations, we can obtain the final closed-form expression of (\ref{eq:12}) as shown in (\ref{eq:17}).

\section{Derivation of Useful Function} \label{appendix_4}
In this appendix, we derive (\ref{eq:24}) by deriving the special case from $n_1=1$ and $n_2=2$ and then we extend this result to the general case for arbitrary $n_1$ and $n_2$.
At first, let $n_1=1$ and $n_2=2$, then we can write (\ref{eq:24}) as the following summation expression
\begin{equation} \small \label{eq:AP_A_1}
\begin{aligned}
&\prod\limits_{j = 1}^{2} {\left( 1 - \exp \left( { - a_{i_j}} \right) \right)}  = \left( 1 - \exp \left( { - a_{i_1}} \right) \right)\left( 1 - \exp \left( { - a_{i_2}} \right) \right)
\\
&=1- \exp \left( { - a_{i_1}} \right)-\exp \left( { - a_{i_2}} \right)+\exp \left( { - a_{i_1}}{ - a_{i_2}} \right).
\end{aligned}
\end{equation}
Similarly, for $n_1=1$ and $n_2=3$, we can write (\ref{eq:24}) as
\begin{equation} \small \label{eq:AP_A_2}
\begin{aligned}
&\prod\limits_{j = 1}^{3} {\left( 1 - \exp \left( { - a_{i_j}} \right) \right)}  = \left( 1 - \exp \left( { - a_{i_1}} \right) \right)\left( 1 - \exp \left( { - a_{i_2}} \right) \right)\left( 1 - \exp \left( { - a_{i_3}} \right) \right)
\\
&=1- \exp \left( { - a_{i_1}} \right)-\exp \left( { - a_{i_2}} \right) -\exp \left( { - a_{i_3}} \right) + \exp \left( { - a_{i_1}} { - a_{i_2}} \right) + \exp \left( { - a_{i_1}} { - a_{i_3}} \right)
\\
&\quad + \exp \left( { - a_{i_2}}  { - a_{i_3}} \right)- \exp \left( { - a_{i_1}} { - a_{i_2}} { - a_{i_3}} \right).
\end{aligned}
\end{equation}
After simplification with a few manipulations, we can re-write the multiple product expressions in (\ref{eq:AP_A_1}) and (\ref{eq:AP_A_2})  as the following simplified summation expressions, respectively
\begin{equation} \small \label{eq:AP_A_3}
\begin{aligned}
&\prod\limits_{j = 1}^{2} {\left( 1 - \exp \left( { - a_{i_j}} \right) \right)}
\\
&=1 + \left( { - 1} \right)\sum\limits_{{{j'}_1} = 1}^2 {exp\left( { - {a_{{i_{{{j'}_1}}}}}} \right)}  + {\left( { - 1} \right)^2}\sum\limits_{{{j'}_1} = 1}^1 {\sum\limits_{{{j'}_2} = {{j'}_1} + 1}^2 {\exp \left( { - {a_{{i_{{{j'}_1}}}}} - {a_{{i_{{{j'}_2}}}}}} \right)} },
\end{aligned}
\end{equation}
and
\begin{equation} \small \label{eq:AP_A_4}
\begin{aligned}
&\prod\limits_{j = 1}^{3} {\left( 1 - \exp \left( { - a_{i_j}} \right) \right)}
\\
&=1 + \left( { - 1} \right)\sum\limits_{{{j'}_1} = 1}^3 {exp\left( { - {a_{{i_{{{j'}_1}}}}}} \right)}  + {\left( { - 1} \right)^2}\sum\limits_{{{j'}_1} = 1}^2 {\sum\limits_{{{j'}_2} = {{j'}_1} + 1}^3 {\exp \left( { - {a_{{i_{{{j'}_1}}}}} - {a_{{i_{{{j'}_2}}}}}} \right)} }
\\
&\quad+ {\left( { - 1} \right)^3}\sum\limits_{{{j'}_1} = 1}^1 {\sum\limits_{{{j'}_2} = {{j'}_1} + 1}^2 {\sum\limits_{{{j'}_3} = {{j'}_2} + 1}^3 {\exp \left( { - {a_{{i_{{{j'}_1}}}}} - {a_{{i_{{{j'}_2}}}}} - {a_{{i_{{{j'}_3}}}}}} \right)} } }.
\end{aligned}
\end{equation}
As results, after simplifying and generalizing the above equations, we can obtain the generalized expression of (\ref{eq:AP_A_1}) and (\ref{eq:AP_A_2}) with arbitrary $n_1$ and $n_2$ as
\begin{equation} \small \label{eq:AP_A_5}
\prod\limits_{j = n_1}^{n_2} {\left( 1 - \exp \left( { - a_{i_j}} \right) \right)}  = 1 + \sum\limits_{l = 1}^{n_2 - {n_1}+1} {{{\left( { - 1} \right)}^l}\sum\limits_{{{j'}_1} = {{j'}_0} +n_1}^{n_2 - l + 1} { \cdots \sum\limits_{{{j'}_l} = {{j'}_{l - 1}} + 1}^{n_2} {\exp \left( { - \sum\limits_{m = 1}^l {a_{i_{{j'}_m} }}} \right)} } },
\end{equation}
where ${j'}_0=0$.

\bibliographystyle{ieeetran}
\bibliography{IEEEabrv,thesis}

\begin{thebibliography}{10}
\providecommand{\url}[1]{#1}
\csname url@rmstyle\endcsname
\providecommand{\newblock}{\relax}
\providecommand{\bibinfo}[2]{#2}
\providecommand\BIBentrySTDinterwordspacing{\spaceskip=0pt\relax}
\providecommand\BIBentryALTinterwordstretchfactor{4}
\providecommand\BIBentryALTinterwordspacing{\spaceskip=\fontdimen2\font plus
\BIBentryALTinterwordstretchfactor\fontdimen3\font minus
  \fontdimen4\font\relax}
\providecommand\BIBforeignlanguage[2]{{%
\expandafter\ifx\csname l@#1\endcsname\relax
\typeout{** WARNING: IEEEtran.bst: No hyphenation pattern has been}%
\typeout{** loaded for the language `#1'. Using the pattern for}%
\typeout{** the default language instead.}%
\else
\language=\csname l@#1\endcsname
\fi
#2}}

\bibitem{kn:S_Choi_2008_4}
S.~Choi, M.-S. Alouini, K.~A. Qaraqe, and H.-C. Yang, ``Finger replacement
  method for {RAKE} receivers in the soft handover region,'' \emph{{IEEE}
  Trans. Wireless Commun.}, vol.~7, no.~4, pp. 1152--1156, Apr. 2008.

\bibitem{kn:S_Choi_2008_1}
------, ``Finger replacement schemes for {RAKE} receivers in the soft handover
  region with multiple base stations,'' \emph{{IEEE} Trans. Veh. Technol.},
  vol.~57, no.~4, pp. 2114--2122, July 2008.

\bibitem{kn:S_Choi_2008_2}
------, ``Soft handover overhead reduction by {RAKE} reception with finger
  reassignment,'' \emph{{IEEE} Trans. Commun.}, vol.~56, no.~2, pp. 213--221,
  Feb. 2008.

\bibitem{kn:S_Choi_2008_3}
------, ``Finger assignment schemes for {RAKE} receivers with multiple-way soft
  handover,'' \emph{{IEEE} Trans. Wireless Commun.}, vol.~7, no.~2, pp.
  495--499, Feb. 2008.

\bibitem{kn:MS_GSC}
H.-C. Yang, ``New results on ordered statistics and analysis of
  minimum-selection generalized selection combining {(GSC)},'' \emph{{IEEE}
  Trans. Wireless Commun.}, vol.~5, no.~7, pp. 1876--1885, July 2006.

\bibitem{kn:alouini_wi_j3}
{M.-S. Alouini} and M.~K. Simon, ``An {MGF}-based performance analysis of
  generalized selective combining over {R}ayleigh fading channels,''
  \emph{{IEEE} Trans. Commun.}, vol.~48, no.~3, pp. 401--415, Mar. 2000.

\bibitem{kn:unified_approach}
S.~S. Nam, M.-S. Alouini, and H.-C. Yang, ``A {MGF}-based unified framework to
  determine the joint statistics of partial sums of ordered random variables,''
  \emph{{IEEE} Trans. Inform. Theory}, vol.~56, no.~8, pp. 5655--5672, Nov.
  2010.

\bibitem{kn:sungsiknam2013_ISIT}
S.~S. Nam, H.-C. Yang, M.-S. Alouini, and D.~I. Kim, ``Exact capture
  probability analysis of {GSC} receivers over i.n.d. {Rayleigh} fading
  channels,'' in \emph{Proc. of {IEEE} International Symposium on Information
  Theory and its Applications (ISITA'08)}, Istanbul, Turkey, July 2013, pp.
  56--60.

\bibitem{kn:IND_MGF_sungsiknam_1}
S.~S. Nam, M.-S. Alouini, H.-C. Yang, and D.~I. Kim, ``An {MGF}-based unified
  framework to determine the joint statistics of partial sums of ordered i.n.d.
  random variables,'' \emph{{IEEE} Trans. Signal Processing}, vol.~62, no.~16,
  pp. 4270--4283, May 2014.

\bibitem{kn:abramowitz}
M.~Abramowitz and I.~A. Stegun, \emph{Handbook of Mathematical Functions with
  Formulas, Graphs, and Mathematical Tables}, 9th~ed.\hskip 1em plus 0.5em
  minus 0.4em\relax New York, NY: Dover Publications, 1970.

\end{thebibliography}

\clearpage
\begin{figure}
\centering
\includegraphics[width=5in]{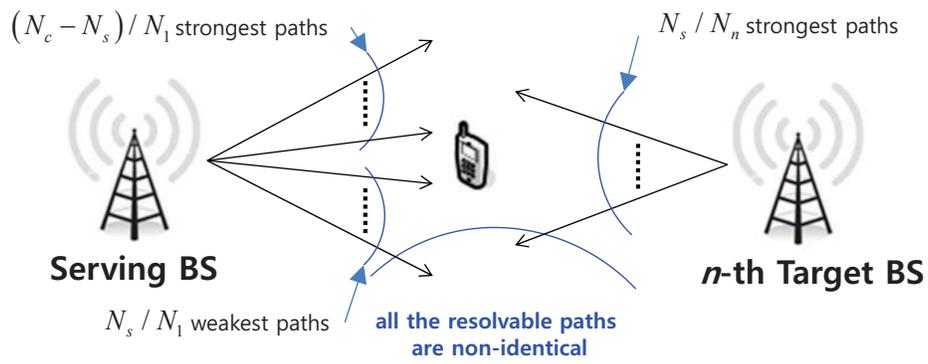}
\caption{Finger replacement schemes for RAKE receivers in the soft handover region.}
\label{example_2}
\end{figure}

\newpage
\clearpage

\begin{figure}
\centering
\includegraphics[width=4.8in]{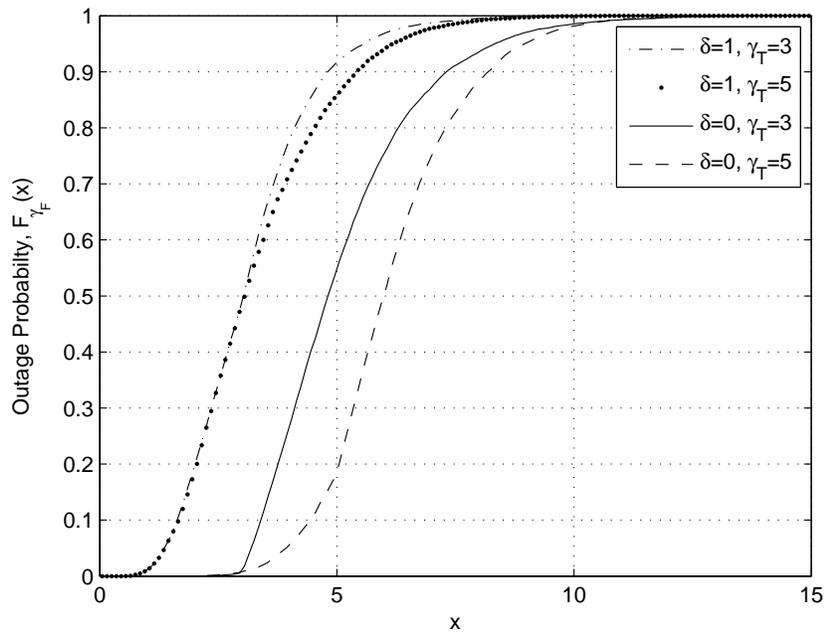}
\centerline{ (a) $N_s=1$ }
\includegraphics[width=4.8in]{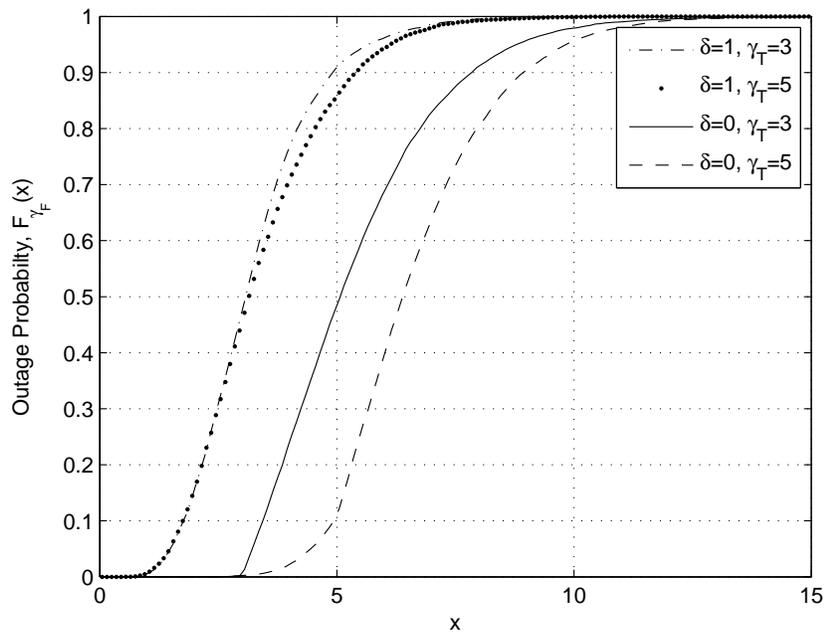}
\centerline{ (b) $N_s=2$ } 
\caption{Outage probability of finger
replacement schemes for RAKE receivers in the soft handover region
over i.n.d. Rayleigh fading channels when $L=4$, $N_1=\cdots=N_4
=5$, and $N_c=3$.} \label{results}
\end{figure}

\end{document}